\documentclass[10pt,twocolumn,showpacs,amsmath,amssymb,floatfix,superscriptaddress]{revtex4-2}

\usepackage{graphicx}
\usepackage{dcolumn}
\usepackage{bm}
\usepackage{color}
\usepackage{txfonts}
\usepackage{microtype}
\usepackage{hyperref}
\usepackage[english]{babel}
\usepackage{slashed}
\usepackage{gensymb}
\usepackage{epsfig}
\usepackage[normalem]{ulem}
\usepackage{amsmath} 
\usepackage{array}
\usepackage{booktabs}
\allowdisplaybreaks[4]

\begin{document}

\renewcommand{\figurename}{FIG}	

\title{ Nuclear Excitation and Control Induced by Intense Vortex Laser}

\author{Zhi-Wei Lu}
\thanks{These authors have contributed equally to this work.}
\affiliation{Ministry of Education Key Laboratory for Nonequilibrium Synthesis and Modulation of Condensed Matter, State key laboratory of electrical insulation and power equipment, Shaanxi Province Key Laboratory of Quantum Information and Quantum Optoelectronic Devices, School of Physics, Xi'an Jiaotong University, Xi'an 710049, China}	
\author{Hanxu Zhang}
\thanks{These authors have contributed equally to this work.}
\affiliation{Graduate School, China Academy of Engineering Physics, Beijing 100193, China}
\author{Tao Li}
\affiliation{Graduate School, China Academy of Engineering Physics, Beijing 100193, China}
\author{Mamutjan Ababekri}
\affiliation{Ministry of Education Key Laboratory for Nonequilibrium Synthesis and Modulation of Condensed Matter, State key laboratory of electrical insulation and power equipment, Shaanxi Province Key Laboratory of Quantum Information and Quantum Optoelectronic Devices, School of Physics, Xi'an Jiaotong University, Xi'an 710049, China}	
\author{Xu Wang}\email{xwang@gscaep.ac.cn}
\affiliation{Graduate School, China Academy of Engineering Physics, Beijing 100193, China}
\affiliation{Southern Center for Nuclear-Science Theory, Institute of Modern Physics, Chinese Academy of Sciences, Huizhou, Guangdong 516000, China}
\author{Jian-Xing Li}\email{jianxing@xjtu.edu.cn}
\affiliation{Ministry of Education Key Laboratory for Nonequilibrium Synthesis and Modulation of Condensed Matter, State key laboratory of electrical insulation and power equipment, Shaanxi Province Key Laboratory of Quantum Information and Quantum Optoelectronic Devices, School of Physics, Xi'an Jiaotong University, Xi'an 710049, China}	
\affiliation{Department of Nuclear Physics, China Institute of Atomic Energy, P.O. Box 275(7), Beijing 102413, China}

	\date{\today}
	
\begin{abstract}
The existing intense laser-based approaches for nuclear excitation offer ultrafast temporal resolution and high efficiency compared to traditional accelerator probes. However, controlling nuclear properties such as spin and magnetic moment remains an unprecedented challenge. Here, we put forward a novel method for nuclear excitation and control induced by intense vortex lasers. We develop a theory incorporating the orbital angular momentum (OAM) of vortex laser within the nuclear hyperfine mixing framework. We find that intense vortex laser can effectively excite hydrogen-like thorium-229 nucleus and induce three-dimensional rotation of the nuclear magnetic moment. This rotation arises from the localized electromagnetic field and new transition channels excited by the vortex laser, and can be reconstructed through radiation spectrum analysis. Moreover, the OAM of vortex laser enables the chaotic system to exhibit topologically protected periodic patterns in nuclear excitation and radiation, facilitating precise experimental measurements. Our findings underscore the potential of vortex laser for high-precision nuclear control and imaging, deepening our understanding of nuclear properties and hyperfine structure, and advancing quantum information and nuclear technologies. 
\end{abstract}

\maketitle

Efficient excitation and control of nuclei have long been pursued in nuclear physics, significantly impacting various applications, including quantum information (enhancing qubit coherence) \cite{fuchs2011quantum,zaiser2016enhancing}, medical imaging (improving magnetic resonance imaging resolution) \cite{andrew1958nuclear,pykett1982principles,lambert2019nuclear}, materials science (developing novel magnetic materials) \cite{coronado2020molecular,gao2022nuclear}, and fundamental physics (investigating nuclear structure and interactions) \cite{peik2003nuclear,tiedau2024laser,walker1999energy,chiara2018isomer,baldwin1997recoilless,tkalya2011proposal}. 
Traditional probes for nuclear excitation and control primarily include nuclear magnetic resonance probes that combine external magnetic fields and radiofrequency pulses \cite{lambert2019nuclear}, as well as relativistic ($\sim$MeV) particle beams from accelerators, such as $\gamma$ rays \cite{zilges2022photonuclear}, electrons \cite{downer2018diagnostics}, neutrons \cite{dubbers2011neutron}, and ions \cite{shiltsev2021modern}. The former are mainly suitable for light nuclei with a spin angular momentum of $1/2$, with transition energies typically in the microelectronvolt ($\mu$eV) range, limiting their effectiveness for higher-energy transitions and heavy nuclei. The latter is constrained by the limited luminosity of the accelerator beam, which restricts time resolution and excitation efficiency. 
The rapid development of ultraintense, ultrashort laser facilities has achieved peak intensities of approximately $10^{23} \, {\rm W/cm^2}$, with pulse durations in the tens of femtoseconds \cite{balabanski2017new,zhang2020laser,yoon2021realization}. This progress has catalyzed the development of laser-based proposals for nuclear excitation, encompassing both experimental \cite{feng2022femtosecond,feng2024laser,gargiulo2024revisiting,jacob2025enhanced} and theoretical \cite{wang2021exciting,qi2023isomeric,wang2024isomeric} methodologies that exhibit ultrafast time resolution and ultrahigh excitation efficiency. Nevertheless, the excitation cross-sections associated with these approaches remain relatively small. 
Very recently, research \cite{zhang2024highly} predicts that the highly nonlinear interaction between intense lasers and hydrogenlike thorium-229 ions ($^{229}{\rm Th}^{89+}$) can excite over 10\% of these ions into the isomeric state with a single femtosecond laser pulse, accompanied by high-order harmonic emission. In this scenario, with one electron outside the nucleus, the strong electromagnetic field generated by electrons near the nucleus induces nuclear hyperfine mixing (NHM) among states with the same total angular momentum, substantially shortening the isomeric lifetime by several orders of magnitude and causing slight shifts in hyperfine energy levels \cite{wycech1993predictions,karpeshin1998rates,shabaev2022ground,wang2024substantial}. Unfortunately, this highly nonlinear interaction renders the excitation and radiation patterns of the system chaotic and unpredictable, limiting control over the nucleus.
Overall, current researches are still focused on effectively exciting nuclei, and controlling properties such as nuclear spin and nuclear magnetic moment presents unprecedented challenges.

  \begin{figure}[!t]	
 	\setlength{\abovecaptionskip}{0.cm}
 	\setlength{\belowcaptionskip}{-0.cm}
 	\centering\includegraphics[width=1\linewidth]{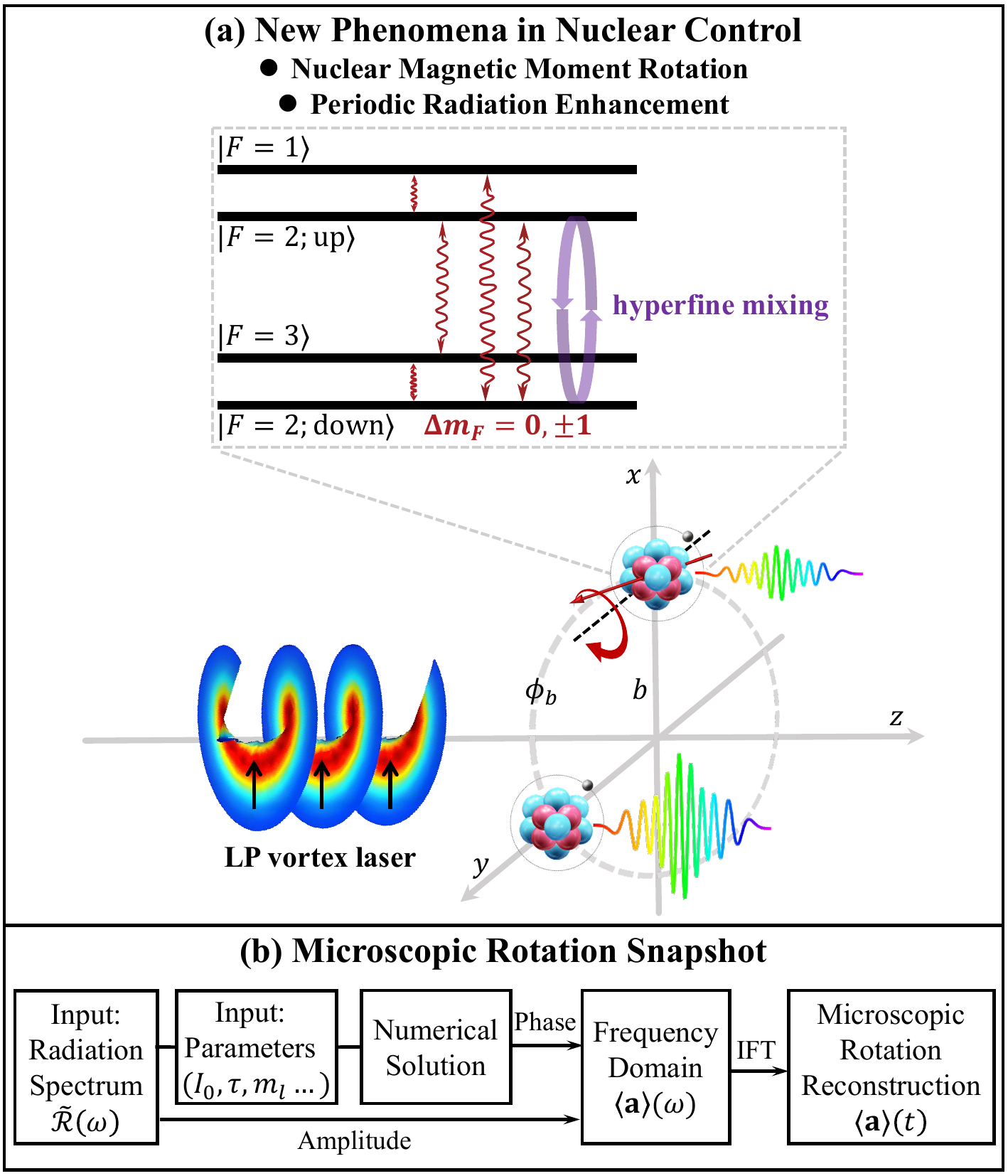}
 	\vspace{-0.8 cm}
 	\begin{picture}(300,25)
 		
 	\end{picture}
 	\caption{ Interaction scenario between an intense vortex laser and the $^{229}{\rm Th}^{89+}$ ion. A linearly polarized (LP) vortex laser in the $x$-direction (black arrow) propagates along the $z$-axis. The ion's position relative to the vortex laser's central axis is defined by the impact parameter $\textbf{b}$, comprising distance $b$ and azimuthal angle $\phi_b$.  	State mixing of the two levels with total angular momentum $F=2$ arises from the NHM effect. The transition selection rules for changes in the magnetic quantum number $\Delta m_F$ are modified from the previous $\pm1$ to $0,\pm1$.
 	(a) New Phenomena in Nuclear Control: Vortex laser-induced high-order harmonic radiation (rainbow-colored pulses) exhibits periodic enhancement at various azimuthal angles $\phi_b$, with the period determined by OAM projection $m_l$. The average acceleration $\langle \textbf{a}\rangle$ of the nuclear magnetic moment rotates during radiation (red arrow). 
 	(b) Microscopic Rotation Snapshot: The radiation spectrum $\tilde{\mathcal{R}}(\omega)$, combined with theoretical calculations, reconstructs the acceleration's frequency-domain signal. Here $\omega$, $I_0$ and $\tau$ denote radiation frequency, laser peak intensity, and pulse duration, respectively. Inverse Fourier transform (IFT) yields the time-domain rotational signal. }
 	\label{fig1}
 \end{figure}

Meanwhile, recent advancements in the fabrication of phase masks with nanometer precision has rendered it possible to control the coherent superposition of matter waves, which produces typical interference patterns through spatial wave function reshaping \cite{lee2019laguerre,luski2021vortex,clark2015controlling,mcmorran2011electron}. Particularly interesting are vortex photons, described by wave functions with helical phases that carry intrinsic orbital angular momentum (OAM) along their propagation axis \cite{allen1992orbital,knyazev2018beams}. Currently, vortex photons ranging from visible to x‐ray (eV$\sim$keV) have been experimentally generated via optical mode conversion, high harmonic techniques, and coherent radiation in helical undulators and contemporary laser facilities \cite{shen2019optical,peele2002observation,terhalle2011generation,gariepy2014creating,hemsing2013coherent}. Vortex photons, arising from the new degree of freedom, provide unique advantages in optical manipulation, quantum information, and imaging techniques \cite{shen2019optical,ivanov2022promises}, which have been predicted and demonstrated in atomic \cite{lange2022excitation,afanasev2018experimental,schmiegelow2016transfer,das2024high,shi2024advances}, molecular \cite{forbes2018optical,trawi2023molecular}, and larger scales systems \cite{garces2003observation,swartzlander2008astronomical}. Additionally, our previous theoretical studies on relativistic vortex particles, such as $\gamma$ rays \cite{lu2023manipulation} and electrons \cite{lu2025angular}, interacting with nuclei in the giant resonance regime (tens of MeV), indicate that vortex particles have the potential to manipulate nuclear transitions and provide new insights into nuclear structure. 
These findings motivate our ongoing exploration of the interactions between contemporary intense vortex lasers and nuclei exhibiting the NHM effect, which raises questions about the phenomena of nuclear excitation and radiation that may emerge, as well as whether vortex lasers can be utilized for nuclear control.

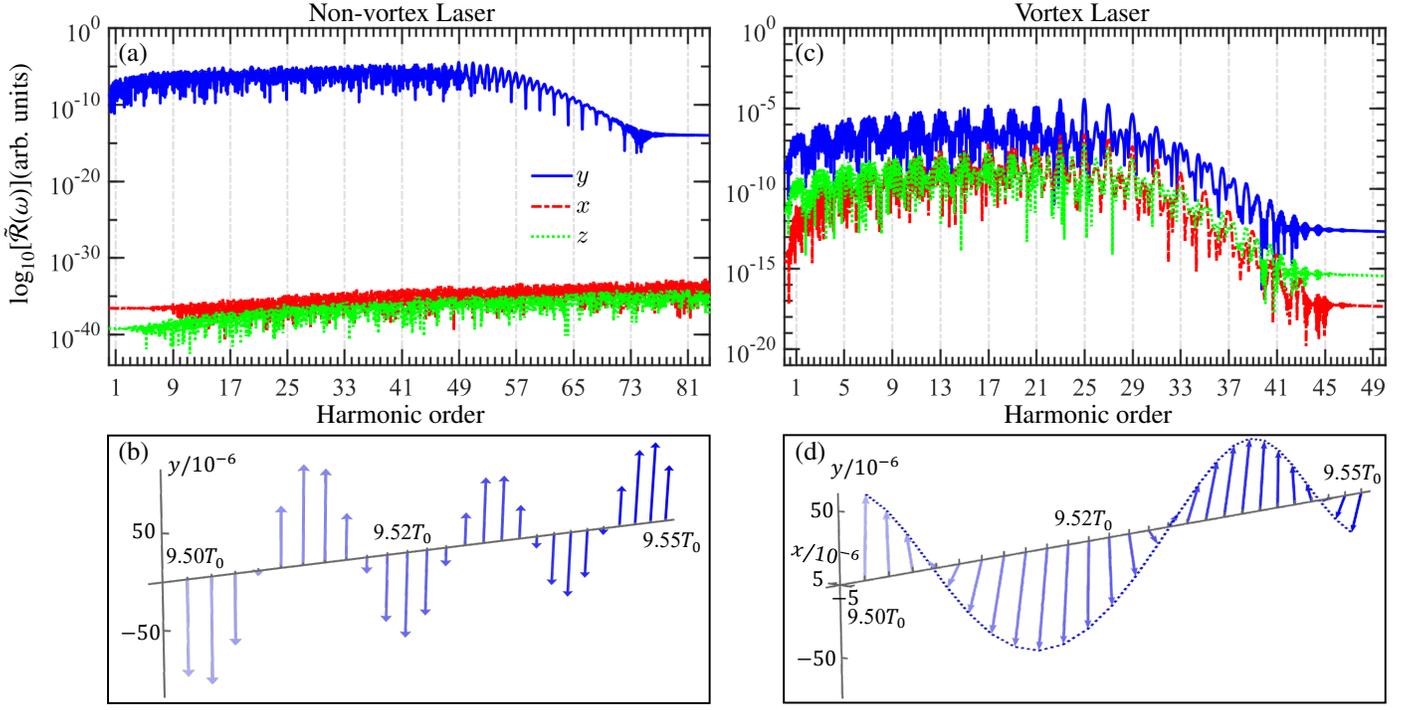
\begin{figure*}[!t]	
	\setlength{\abovecaptionskip}{0.cm}
	\setlength{\belowcaptionskip}{-0.cm}
	\centering\includegraphics[width=1\linewidth]{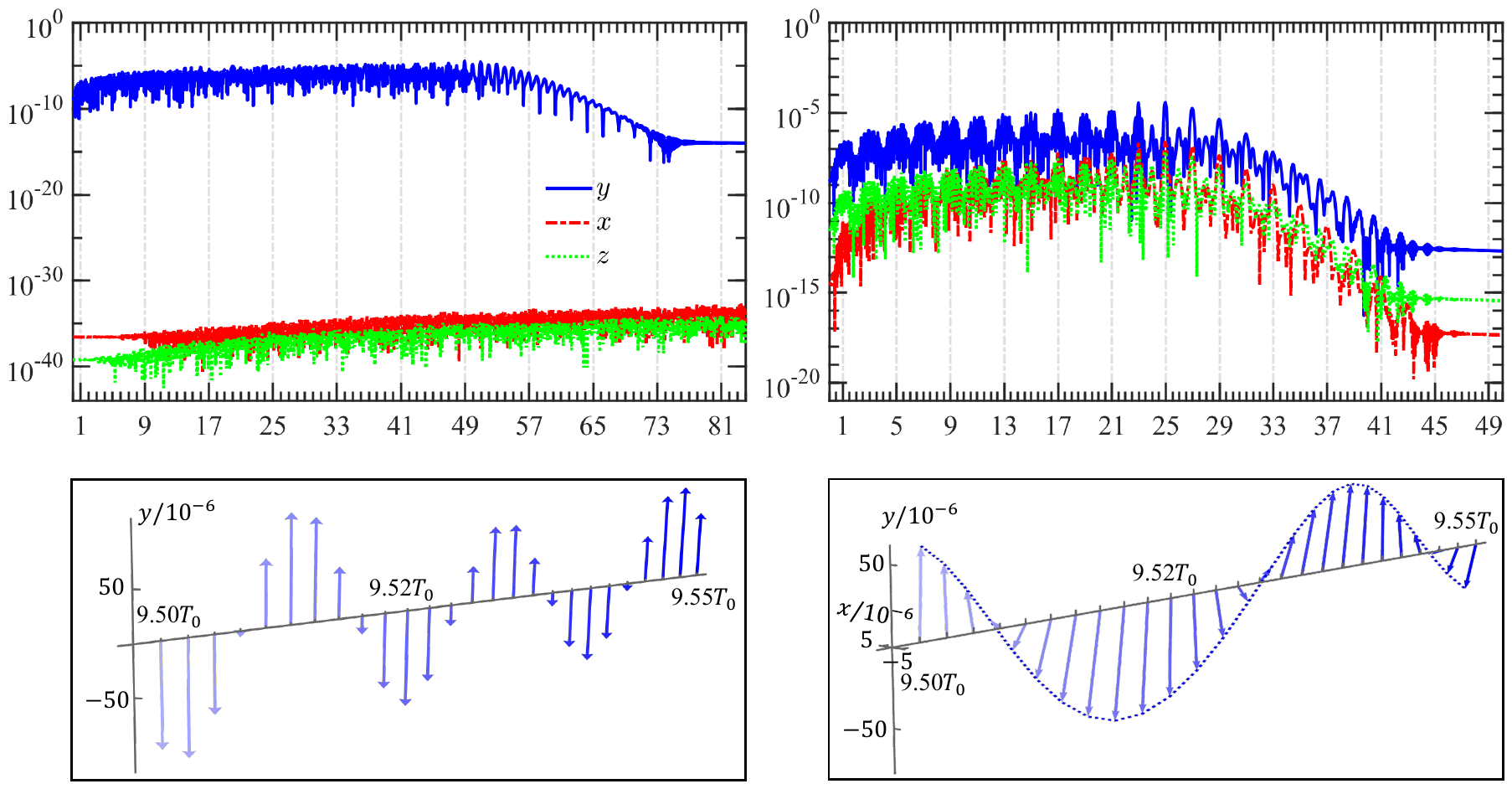}
	\vspace{-0.9 cm}
	\begin{picture}(300,25)
		\put(-77,272){\normalsize{(a)}}
		\put(-5,287){\normalsize{Non-vortex Laser}}
		\put(179,272){\normalsize{(c)}}
		\put(-77,121){\normalsize{(b)}}
		\put(262,287){\normalsize{Vortex Laser}}
		\put(179,121){\normalsize{(d)}}
		\put(-120,180){\rotatebox{90}{\normalsize{${\rm log}_{10}[\tilde{\mathcal{R}}(\omega)]$(arb. units)}}} 
		\put(-2,135){\normalsize{Harmonic order}}
		\put(258,135){\normalsize{Harmonic order}}
	\end{picture}
	\caption{ (a) Radiation spectrum of the non-vortex laser-driven $^{229}{\rm Th}^{89+}$ system, with peak intensity of $I_0=10^{22}$ W/cm$^2$ and pulse duration of 20 cycles ($\tau=20 T_0$). Lines in three colors (styles) represent the contributions from $\tilde{\mathcal{R}}_{x}$, $\tilde{{\mathcal{R}}}_{y}$, and $\tilde{\mathcal{R}}_{z}$. The horizontal axis is in units of laser photon energy (1.55 eV). (b) Time evolution of the reconstructed $\langle{\textbf{a}}\rangle$ (atomic units) between 9.5$T_0$ and 9.55$T_0$, with arrow color deepening to indicate time passage; arrows are spaced by 0.002$T_0$. (c) and (d) are similar to (a) and (b), respectively, but for a vortex laser. Parameters: OAM projection $m_l=1$, pitch angle $\theta_k=5^\circ$, beam waist $w_0=10 \lambda_0$, impact parameter $\varkappa b=2$ and $\phi_b=0^\circ$.}
	\label{fig2}
\end{figure*}

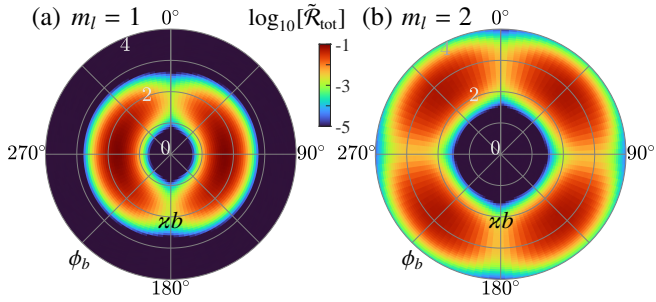
\begin{figure}[!t]	
	\setlength{\abovecaptionskip}{0.cm}
	\setlength{\belowcaptionskip}{-0.cm}
	\centering\includegraphics[width=1\linewidth]{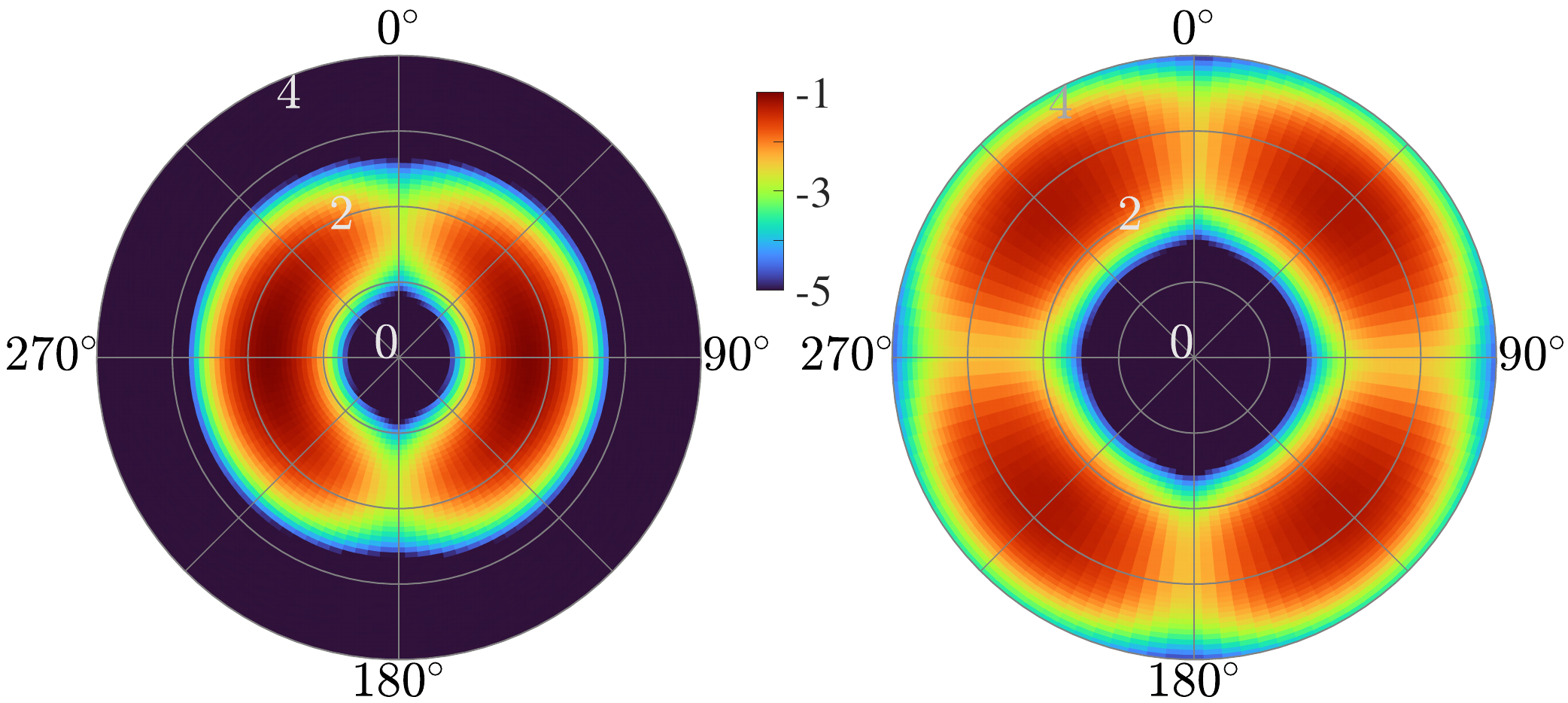}
	\vspace{-0.8 cm}
	\begin{picture}(300,25)
		\put(10,129){\normalsize{(a)}}
		\put(25,129){\normalsize{$m_l=1$}}
		\put(135,129){\normalsize{(b)}}
		\put(150,129){\normalsize{$m_l=2$}}
		\put(92,129){\small{${\rm log}_{10}[\tilde{\mathcal{R}}_{\rm tot}]$}}
		\put(24,38){\small{$\phi_b$}}
		\put(150,38){\small{$\phi_b$}}
		\put(58,52){\small{$\varkappa b$}}
		\put(183,52){\small{$\varkappa b$}}
	\end{picture}
	\caption{(a) Distribution of radiation spectrum ${\rm log}_{10}[\tilde{\mathcal{R}}_{\rm tot}]$ on the impact parameter $\textbf{{b}}$ for the vortex laser-driven $^{229}{\rm Th}^{89+}$ system, where $\tilde{\mathcal{R}}_{\rm tot}=\tilde{\mathcal{R}}_{x}+\tilde{\mathcal{R}}_{y}+\tilde{\mathcal{R}}_{z}$. The radial component is $\varkappa b$, and the azimuthal component is $\phi_b$. Results correspond to the 27th harmonic order with laser parameters similar to those in Fig. \ref{fig2}(c). (b) Similar to (a), but for $m_l=2$ and the 17th harmonic order.  }
	\label{fig3}
\end{figure}

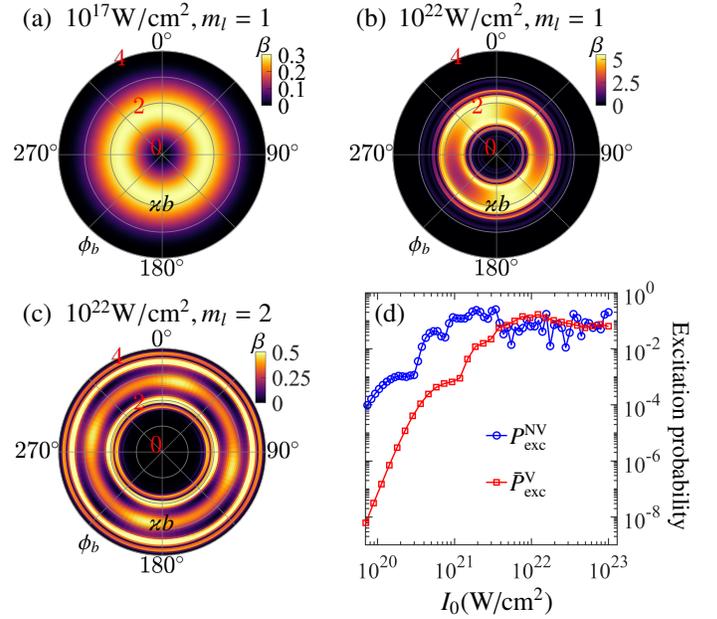
\begin{figure}[!t]	
	\setlength{\abovecaptionskip}{0.cm}
	\setlength{\belowcaptionskip}{-0.cm}
	\centering\includegraphics[width=1\linewidth]{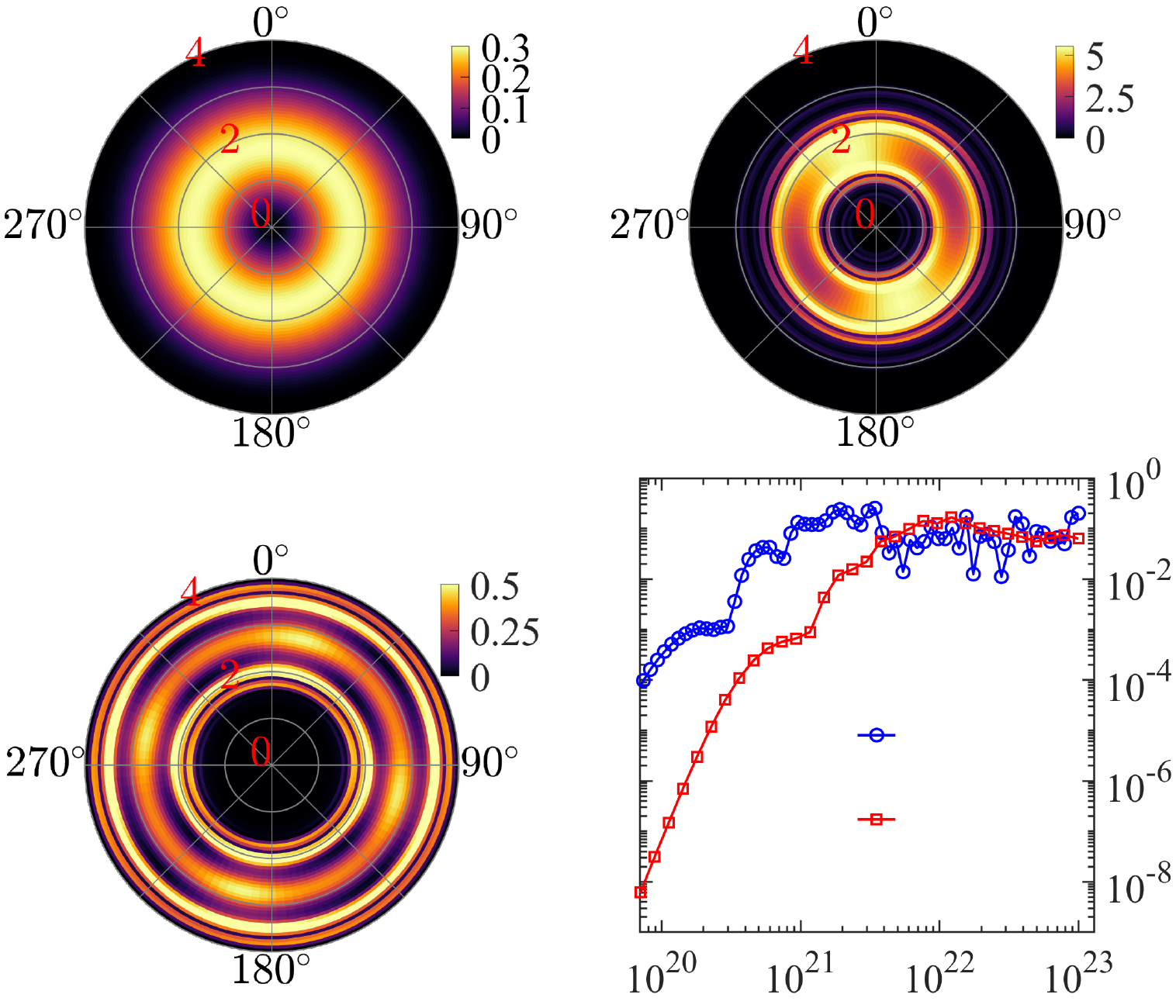}
	\vspace{-0.4 cm}
	\begin{picture}(300,25)
		\put(4,238){\normalsize{(a)}}
		\put(20,238){\normalsize{$10^{17}{\rm W/cm}^2,m_l=1$}}
		\put(130,238){\normalsize{(b)}}
		\put(146,238){\normalsize{$10^{22}{\rm W/cm}^2,m_l=1$}}
		\put(4,126){\normalsize{(c)}}
		\put(20,126){\normalsize{$10^{22}{\rm W/cm}^2,m_l=2$}}
		\put(137,126){\normalsize{(d)}}
		\put(93,228){\small{$\beta$}}
		\put(219,228){\small{$\beta$}}
		\put(91,116){\small{$\beta$}}
		\put(162,17){\normalsize{$I_0$(W/cm$^2$)}}
		\put(188,80){\small{$P^{\rm NV}_{\rm exc}$}}
		\put(188,62){\small{$\bar{P}^{\rm V}_{\rm exc}$}}
		\put(248,126){\rotatebox{270}{\small{Excitation probability}}}
		\put(25,153){\small{$\phi_b$}}
		\put(150,153){\small{$\phi_b$}}
		\put(25,40){\small{$\phi_b$}}
		\put(52,168){\small{$\varkappa b$}}
		\put(179,168){\small{$\varkappa b$}}
		\put(52,47){\small{$\varkappa b$}}
	\end{picture}
	\caption{Nuclear excitation probability of $^{229}{\rm Th}^{89+}$ from the ground state to the isomeric state $|F=2;{\rm up}\rangle$ at the last moment following the completion of the laser pulse. (a)-(c) Distribution of excitation probability ratio ($\beta=P^{\rm V}_{\rm exc}/P^{\rm NV}_{\rm exc}$) between vortex and non-vortex lasers on the impact parameter $\textbf{{b}}$ for various $I_0$ and $m_l$. (d) Excitation probability versus $I_0$ with pulse duration of $\tau=10T_0$. The average excitation probability for the vortex laser, $\bar{P}^{\rm V}_{\rm exc}$, is calculated over $\textbf{b}\leq \textbf{b}_0$ as $\bar{P}^{\rm V}_{\rm exc}=\int_0^{\textbf{b}_0} P^{\rm V}_{\rm exc}(\textbf{b}) \frac{d^2 \textbf{b}}{\pi \textbf{b}_0^2}$, where $|\textbf{b}_0|=3/\varkappa$. Other laser parameters match those in Fig. \ref{fig2}(c).   }
	\label{fig4}
\end{figure}

In this Letter, we put forward a novel method for nuclear excitation and control using intense vortex laser. We develop a theory incorporating the OAM of vortex laser within the NHM framework. We find that intense vortex laser induces two interesting phenomena in nuclear control [Fig. \ref{fig1}(a)]. (1) Nuclear Magnetic Moment Rotation: The local electromagnetic field of the vortex laser and the reconfiguration of the system's population by the newly excited transition channels result in a three-dimensional rotation of the nuclear magnetic moment. The radiation spectra retain signatures of this rotational effect, enabling the microscopic rotation reconstruction [Fig. \ref{fig1}(b)], which enhances our understanding of fundamental nuclear properties (e.g., magnetic susceptibility and spin dynamics) and nuclear structures (e.g., hyperfine energy level structures and populations). Notably, this rotational effect is distinct from nuclear rotational excitation \cite{bohr1976rotational} (details in Fig. \ref{fig2}). (2) Periodic Radiation Enhancement: As the laser intensity increases, we observe periodic enhancements in the radiation spectra at various azimuthal angles determined by the OAM of vortex laser (details in Fig. \ref{fig3}). The peak values of radiation spectra can be significantly amplified by several orders of magnitude compared to that in non-vortex case. Correspondingly, the isomeric excitation probability, while exhibiting some randomness, also displays a periodic pattern (details in Fig. \ref{fig4}). This indicates that introducing OAM into an otherwise chaotic system leads to the emergence of topologically protected features, presenting ordered and regular patterns that facilitate experimental measurement. Additionally, position-dependence offers a new dimension for precise nuclear state control.

The traditional photonuclear interaction theory \cite{eisenberg1976nuclear} is inadequate for vortex light with OAM. Recent advancements in vortex light-nucleus interaction theory have either focused on bare nuclei with only nuclear spin \cite{lu2023manipulation} or overlooked NHM effects in ions \cite{kirschbaum2024photoexcitation}. We develop a theory that self-consistently incorporates the OAM of vortex laser in NHM framework. 
Specifically, using the $^{229}{\rm Th}^{89+}$ ion as an example, the Hamiltonian of the laser-nucleus-electron system can be expressed as $H=H_0+H_{\rm I}(t) = H_{\rm e}+H_{\rm n} +V_{\rm HF} +H_{\rm I}(t)$, where $V_{\rm HF}$ represents the hyperfine interaction \cite{alder1956study,cowan1981theory}, causing minor adjustments in the energies of the hyperfine levels and state mixing between two $F=2$ levels, with the NHM coefficient $c_m\approx-0.031$ in our calculations (details in \cite{supplemental}). The interaction Hamiltonian with the laser, $H_{\rm I}(t)$, includes the contributions from the current density operators of the electron and nucleus:
\begin{eqnarray}
	H_{\rm I}(t) = -\frac{1}{c} \int [ \boldsymbol{j}_{\rm e}(\boldsymbol{r})+ \boldsymbol{j}_{\rm n}(\boldsymbol{r}) ] \cdot \boldsymbol{A}^{\rm V}(\boldsymbol{r},t) d\boldsymbol{r}, 
\end{eqnarray}
where $\boldsymbol{A}^{\rm V}$ is the vector potential of vortex laser, exemplified by the Bessel-Gauss mode \cite{sheppard1978gaussian} in our calculations, with beam waist $w_0$. Different vortex modes (e.g., Bessel and Laguerre-Gaussian) \cite{durnin1987exact,afanasev2018experimental} exhibit varying asymptotic behaviors far from the beam axis, but this does not affect the qualitative conclusions (details in \cite{supplemental}). The vortex laser is assumed to be LP in the $x$-direction under the paraxial approximation, where the transverse momentum of the photon is much smaller than its longitudinal counterpart, $\varkappa=|\boldsymbol{k}_\perp|\ll k_z$ \cite{schulz2020generalized}. The laser wavelength is $\lambda_0=800$ nm and the temporal envelope is $f(t)=\sin^2(\pi t/\tau)$, with $\tau=NT_0$ as the pulse duration and $T_0=2.67$ fs as the optical period. Considering the transition between $|F=2;{\rm down}\rangle$ and $|F=2;{\rm up}\rangle$, with the magnetic quantum numbers of the two states being $\{m_{\rm down},m_{\rm up}\}$. The time-independent interaction energies $E^{\rm V}_{\rm I}$ in the interaction Hamiltonian can be expressed as 
\begin{eqnarray}
	E^{\rm V}_{\rm I} &\approx& 
	\begin{pmatrix}  
		2 & 1 & 2 \\
		m_{\rm down} & \Delta m_F & -m_{\rm up} 
	\end{pmatrix} \mathcal{E}_0 
	\left[   \sqrt{B(M1)} + c_m   \mu_e    \right] \nonumber\\
	&&* e^{-b^2/w_0^2} e^{i(m_l+\Lambda-\Delta m_F)\phi_b} d^1_{\Delta m_F \Lambda}(\theta_k) J_{m_l+\Lambda-\Delta m_F}(\varkappa b)  .
\end{eqnarray}
The detailed expression is provided in \cite{supplemental}. Here, $\Lambda$, $\mathcal{E}_0$ and $d^1_{\Delta m_F \Lambda}(\theta_k)$ are the helicity, the laser electric field strength and Wigner $d$ function at pitch angle $\theta_k=\varkappa/k_z$, respectively. $B(M1)$ is the reduced magnetic dipole transition probability of the bare nucleus \cite{safronova2013magnetic,thielking2018laser,minkov2021th}, and $\mu_e$ is the magnetic dipole moment of the electron \cite{shabaev2022ground}. We also calculate the LP non-vortex laser with a transverse uniform distribution as in \cite{zhang2024highly}. Note that the interaction energy $E^{\rm V}_{\rm I}$ exhibits two notable features. First, the transition selection rule is modified, similar to the bare nucleus case \cite{lu2023manipulation} but extended to total angular momentum projection. When $b\neq 0$, $\Delta m_F$ can take arbitrary values (i.e., $0,\pm1$ for $M1$ transition). Second, the impact parameter $\mathbf{b}$ introduces an additional degree of nuclear control, influenced by the Bessel-Gauss beam flux and the azimuthal angle $\phi_b$.

The state of the four-level system $| \Psi(t) \rangle$ is expressed in terms of the eigenstates of $H_{\rm 0}$: $|\Psi(t) \rangle = \sum_n C_n(t) e^{-i\omega_n t} |\psi_n\rangle$, where $\{ |\psi_n\rangle \} = \{|F=2;{\rm down}\rangle, |F=3\rangle, |F=2;{\rm up}\rangle, |F=1\rangle  \}$ and the energies of the levels are denoted by $\hbar \omega_n$. The ion initially occupies its ground state $|F=2;{\rm down}\rangle$. The system's evolution is determined by the time-dependent Schrödinger equation \cite{verner2010numerically}, allowing the population of each state $|C_n(t)|^2$ for both non-vortex and vortex laser cases can be obtained at each time step during the laser pulse. The radiation source is the laser-induced magnetic dipole moment, expressed as $\textbf{m}(t)= \langle \Psi(t) | \hat{\textbf{m}}|\Psi(t) \rangle$. The radiation spectrum is calculated as $\tilde{\mathcal{R}}(\omega) \propto \left| \int dt \, \ddot{\mathbf{m}}(t) e^{i\omega t} \right|^2$, where $\ddot{\mathbf{m}}(t)$ is the second derivative of ${\mathbf{m}}(t)$ over time. The acceleration of nuclear magnetic moment is given by $\langle{\textbf{a}}\rangle(t) \propto \ddot{\mathbf{m}}(t)$ \cite{greiner2011quantum}. Since the magnetic dipole moment operator is $\hat{\textbf{m}}\propto rY_{lm}$ \cite{eisenberg1976nuclear}, with $rY_{10}=\sqrt{\frac{3}{4\pi}}z$ and $rY_{1\pm1}=\mp\sqrt{\frac{3}{8\pi}}[x\pm i y]$, allowing the radiation source to be viewed as contributions from the dipole axes along the $x$, $y$ and $z$ directions, denoted as $\tilde{\mathcal{R}}_{x}$, $\tilde{{\mathcal{R}}}_{y}$, and $\tilde{\mathcal{R}}_{z}$.

{\it Nuclear magnetic moment rotation}—
As shown in Fig. \ref{fig2}(a), high-order harmonic radiation from $^{229}{\rm Th}^{89+}$ ion, induced by a LP non-vortex laser in the $x$-direction, arises from the dipole axis contribution in the $y$-direction. Based on the radiation spectrum, the theoretically reconstructed acceleration $\langle{\textbf{a}}\rangle$ oscillates temporally along the laser's magnetic field direction, which is the $y$-axis [see Fig. \ref{fig2}(b)]. For the radiation spectrum of $^{229}{\rm Th}^{89+}$ ion induced by LP vortex laser, contributions are observed not only along the dipole axis in the $y$-direction but also result in high-order harmonics in the $x$ and $z$ directions [see Fig. \ref{fig2}(c)]. The reconstructed acceleration $\langle{\textbf{a}}\rangle$ exhibits temporal rotation [see Fig. \ref{fig2}(d)], induced by the local electromagnetic field of the vortex laser and the reconfiguration of the system's population by the newly excited transition channels $(\Delta m_F=0)$. Specifically, compared to LP non-vortex laser, the electric field of the LP vortex laser includes $E_x$ and $E_z$ components, while the magnetic field contains $B_y$ and $B_z$ components.
Thus, the Poynting vector $\textbf{S}=\textbf{E}\times\textbf{B}$ generates components in the $x$, $y$, and $z$ directions $(S_x,S_y,S_z)$. We also find that when the new transition channels induced by the vortex laser are manually turned off, the radiation spectrum components $\tilde{\mathcal{R}}_{x}$ and $\tilde{\mathcal{R}}_{z}$ decrease by several orders of magnitude, causing the nuclear magnetic moment to dominate oscillations along the $y$ direction (details in \cite{supplemental}). 
Microscopic rotation becomes more pronounced with increasing laser peak intensity $I_0$ and pitch angle $\theta_k$ (details in \cite{supplemental}). The proposed method for reconstructing microscopic rotation remains robust even under 20\% relative random noise in the radiation spectrum, simulating experimental conditions (details in \cite{supplemental}). This stability arises from the crucial phase information [provided by theoretical calculation, see Fig. \ref{fig1} (b)], which encodes the essential rotational details and remains resilient to noise-induced amplitude fluctuations, thereby ensuring reliable reconstruction \cite{yuan2022nuclear}. The two lowest energy levels of the $^{229}{\rm Th}$ nucleus correspond to the band heads of two rotational bands. The rotation discussed here refers to the nuclear magnetic moment rotation induced by the transition between these two rotational energy levels under the vortex laser pulse, distinct from nuclear vibrational and rotational excitation \cite{bohr1976rotational}. Consequently, OAM of vortex laser reshapes the behavior of nuclear magnetic moments, transforming their motion from unidirectional oscillations to three-dimensional rotations. The extraction of the radiation spectrum offers a unique snapshot of microscopic rotation, yielding richer dynamic information about fundamental nuclear properties and structure, potentially improving measurement precision.

{\it Periodic radiation enhancement}—
Figures \ref{fig3}(a) and (b) show the radiation spectra for the impact parameter $\mathbf{b}$ corresponding to $m_l=1$ and $m_l=2$, respectively. We find a periodic enhancement in the radiation spectrum as a function of the azimuthal angle $\phi_b$, determined by the OAM and primarily arises from the contribution of $\tilde{\mathcal{R}}_{y}$. This phenomenon occurs in every harmonic order of the radiation spectrum, illustrated here with an example from the radiation platform region. Specifically, across the entire azimuthal angle, there are $2m_l$ radiation maxima and minima, with the radiation enhancement quantitatively increasing by several orders of magnitude. In contrast, the radiation in the non-vortex case is $\tilde{\mathcal{R}}_{\rm tot}\sim 10^{-5}$ [see Fig. \ref{fig2}(a)], comparable to the minima observed here, indicating that vortex laser substantially enhances the radiation from $^{229}{\rm Th}^{89+}$ ion. Furthermore, we find that the azimuthal angle of the maxima remains invariant for any laser peak intensity and pulse duration, providing a robust diagnostic signal for the OAM of vortex laser (details in \cite{supplemental}). Therefore, the use of vortex laser has the potential to significantly enhance high-order harmonic signals, thereby improving the precision of nuclear radiation measurements. Additionally, the spatial periodicity of the radiation signals may serve as experimental evidence of the nonlinear effects induced by OAM, while also enhancing the understanding of nuclear transitions.

{\it Probability of nuclear excitation}—
At relatively low intensities of $10^{17}$ W/cm$^2$, the excitation probabilities exhibit a linear relationship with the flux of vortex laser, consistent with the intuition that higher photon flux leads to increased excitation probabilities [see Fig. \ref{fig4}(a)]. In contrast to non-vortex laser, the effective peak intensity experienced by the $^{229}{\rm Th}^{89+}$ ion is reduced when using vortex laser, resulting in suppressed excitation probabilities ($\beta<1$). 
With further increases in laser peak intensity, such as $10^{22}$ W/cm$^2$ [see Fig. \ref{fig4}(b)], the light-nucleus interaction enters a highly nonlinear and nonperturbative regime. This is evidenced by the transformation of the radial impact parameter $b$ distribution from a single peak to a multi-peak structure, along with periodic variations in the azimuthal angle $\phi_b$. At the maxima, excitation probabilities may increase several times compared to the non-vortex laser case. This periodicity is determined by the OAM of vortex laser [see Fig. \ref{fig4} (c)], as the interaction energy $E^{\rm V}_{\rm I} \propto e^{im_l\phi_b}$. 
In the linear perturbative regime, the excitation probability $P^{\rm V}_{\rm exc} \propto |E^{\rm V}_{\rm I}|^2$ shows independence from $\phi_b$. However, as laser intensity increases, nonlinear non-perturbative effects become significant, particularly in their dependence on $\phi_b$. Compared to the radiation spectrum (see Fig. \ref{fig3}), although the maxima and minima of the excitation probability spectrum [see Figs. \ref{fig4} (b) and (c)] vary with changes in laser parameters ($I_0, \tau$) and impact parameters $b$, the periodicity determined by the OAM remains unchanged, specifically $P^{\rm V}_{\rm exc}(\phi_b) = P^{\rm V}_{\rm exc}(\phi_b + 180^\circ/m_l)$. This is because the excitation probability at the last moment of the laser pulse, which may exhibit substantial temporal oscillations in the population of isomeric states but ultimately settles at a low value, resulting in strong randomness (details in \cite{supplemental}). The probabilities of the upper level $|F=1\rangle$ exhibit a similar pattern, as seen in \cite{supplemental}. 
In practice, we also consider the collision of the vortex laser with a target where ions are randomly distributed across the incident beam. For such a macroscopic target we average the excitation probability over the impact parameters $\textbf{b}\leq \textbf{b}_0$. As shown in Fig. \ref{fig4}(d), for the case of non-vortex laser, the excitation probability begins to saturate after the peak intensity reaches $10^{21}$ W/cm$^2$, subsequently entering a plateau oscillation region characterized by strong disorder and unpredictability. In contrast, for the vortex laser, saturation is delayed until approximately $10^{22}$ W/cm$^2$, and the plateau region exhibits stability and predictability. 
Therefore, intense non-vortex lasers induce a nonlinear response in $^{229}{\rm Th}^{89+}$ ions, resulting in a chaotic and disordered state of the system. However, upon introducing OAM of vortex laser, a topologically protected periodic pattern emerges. Additionally, the excitation probability averaged over the impact parameter $\mathbf{b}$ shows greater stability and predictable, providing distinctive experimental signals for effective excitation of isomer states and facilitating precise experimental measurements.

{\it Experimental feasibility}—We sequentially discuss the stability of the $^{229}{\rm Th}^{89+}$ ions in intense laser fields, the resulting differences for the lithium-like $^{229}$Th ions ($^{229}{\rm Th}^{87+}$), and experimental suggestions for validation.
Hydrogen-like $^{229}$Th ions remain stable within the intensity range of intense laser fields discussed here. The probability of the $1s_{1/2}$ electron being excited to the $2p_{1/2}$ state (energy gap $\sim93$ keV) is approximately $10^{-5}$ at a laser intensity of $10^{23}$ W/cm$^2$, with even smaller probabilities for excitations to higher states \cite{zhang2024highly}. For the $^{229}{\rm Th}^{87+}$ ions, quantitative differences in excitation and radiation processes primarily arise from the mixing coefficient $c_m$ and the electron magnetic dipole moment $\mu_e$ \cite{shabaev2022ground}. Additionally, because the energy gap between the $2s_{1/2}$ and $2p_{1/2}$ states is only about 245 eV, the oscillation between these two states may introduce an additional layer of control, warranting further exploration and investigation. 
The theoretical findings delineated in this Letter can be implemented using existing experimental configurations, such as ion storage rings \cite{steck2020heavy,ma2015proposal} and Penning traps \cite{ringleb2022high}, which, as reported, have already integrated intense lasers or x-ray free-electron lasers \cite{major2024high,bernitt2012unexpectedly,sturm2019alphatrap,ringleb2022high}. Nonetheless, experiments involving vortex lasers remain scarce. Current PW laser systems, such as PHELIX at GSI \cite{brabetz2015laser} and SULF at SIOM \cite{wang2020hollow}, have already achieved intensities of vortex laser in the $10^{18}$ and $10^{20}$ W/cm$^2$ range \cite{shi2024advances}, with focal spots on the order of 10 $\mu$m. The impact parameter discussed above is likewise on a similar micron scale. We propose potential experimental avenues: for a laser intensity of $10^{18}$ ($10^{20}$) W/cm$^2$, the third (fifth) harmonic exhibits one (two) orders of magnitude of periodic enhancement in the azimuthal radiation spectrum signal (details in \cite{supplemental}). With the ongoing development of 10-100 PW high-power laser systems \cite{radier202210,wang202213}, the analysis and reconstruction of radiation signals induced by stronger laser intensities are expected to provide a snapshot of the microrotation in nuclear magnetic moments.

In conclusion, through an examination of the interaction between intense vortex lasers and $^{229}{\rm Th}^{89+}$ ions, our method could effectively excite and induce three-dimensional rotation of the nuclear magnetic moment on a femtosecond timescale, enabling unprecedented control over its behavior. Furthermore, introducing the OAM of vortex lasers into a chaotic system reveals topologically protected periodic patterns in nuclear excitation and radiation, serving as characteristic experimental signals.
These findings suggest a valuable experimental probe for laser nuclear physics with vortex lasers, creating opportunities for high-precision nuclear control and imaging, and paving the way for innovative applications in fundamental physics, quantum information science, medical imaging, and advanced nuclear technologies, despite the challenges of thorium scarcity.

{\it Acknowledgment}—We thank M. F. Ciappina and B. K. Das for helpful discussions. The work is supported by the National Natural Science Foundation of China (Grants No. 12425510, No. U2267204, No. 12441506, No. 12447106, No. 123B2082, No. 12474484, No. U2330401, No. 12088101), the National Key Research and Development (R$\&$D) Program (Grant No. 2024YFA1610900), the Shaanxi Fundamental Science Research Project for Mathematics and Physics (Grant No. 22JSY014), and the Innovative Scientific Program of CNNC.

\bibliography{ref}

\begin{thebibliography}{80}%
\makeatletter
\providecommand \@ifxundefined [1]{%
 \@ifx{#1\undefined}
}%
\providecommand \@ifnum [1]{%
 \ifnum #1\expandafter \@firstoftwo
 \else \expandafter \@secondoftwo
 \fi
}%
\providecommand \@ifx [1]{%
 \ifx #1\expandafter \@firstoftwo
 \else \expandafter \@secondoftwo
 \fi
}%
\providecommand \natexlab [1]{#1}%
\providecommand \enquote  [1]{``#1''}%
\providecommand \bibnamefont  [1]{#1}%
\providecommand \bibfnamefont [1]{#1}%
\providecommand \citenamefont [1]{#1}%
\providecommand \href@noop [0]{\@secondoftwo}%
\providecommand \href [0]{\begingroup \@sanitize@url \@href}%
\providecommand \@href[1]{\@@startlink{#1}\@@href}%
\providecommand \@@href[1]{\endgroup#1\@@endlink}%
\providecommand \@sanitize@url [0]{\catcode `\\12\catcode `\$12\catcode
  `\&12\catcode `\#12\catcode `\^12\catcode `\_12\catcode `\%12\relax}%
\providecommand \@@startlink[1]{}%
\providecommand \@@endlink[0]{}%
\providecommand \url  [0]{\begingroup\@sanitize@url \@url }%
\providecommand \@url [1]{\endgroup\@href {#1}{\urlprefix }}%
\providecommand \urlprefix  [0]{URL }%
\providecommand \Eprint [0]{\href }%
\providecommand \doibase [0]{https://doi.org/}%
\providecommand \selectlanguage [0]{\@gobble}%
\providecommand \bibinfo  [0]{\@secondoftwo}%
\providecommand \bibfield  [0]{\@secondoftwo}%
\providecommand \translation [1]{[#1]}%
\providecommand \BibitemOpen [0]{}%
\providecommand \bibitemStop [0]{}%
\providecommand \bibitemNoStop [0]{.\EOS\space}%
\providecommand \EOS [0]{\spacefactor3000\relax}%
\providecommand \BibitemShut  [1]{\csname bibitem#1\endcsname}%
\let\auto@bib@innerbib\@empty
\bibitem [{\citenamefont {Fuchs}\ \emph {et~al.}(2011)\citenamefont {Fuchs},
  \citenamefont {Burkard}, \citenamefont {Klimov},\ and\ \citenamefont
  {Awschalom}}]{fuchs2011quantum}%
  \BibitemOpen
  \bibfield  {author} {\bibinfo {author} {\bibfnamefont {G.}~\bibnamefont
  {Fuchs}}, \bibinfo {author} {\bibfnamefont {G.}~\bibnamefont {Burkard}},
  \bibinfo {author} {\bibfnamefont {P.}~\bibnamefont {Klimov}},\ and\ \bibinfo
  {author} {\bibfnamefont {D.}~\bibnamefont {Awschalom}},\ }\bibfield  {title}
  {\bibinfo {title} {A quantum memory intrinsic to single nitrogen--vacancy
  centres in diamond},\ }\href@noop {} {\bibfield  {journal} {\bibinfo
  {journal} {Nat. Phys.}\ }\textbf {\bibinfo {volume} {7}},\ \bibinfo {pages}
  {789} (\bibinfo {year} {2011})}\BibitemShut {NoStop}%
\bibitem [{\citenamefont {Zaiser}\ \emph {et~al.}(2016)\citenamefont {Zaiser},
  \citenamefont {Rendler}, \citenamefont {Jakobi}, \citenamefont {Wolf},
  \citenamefont {Lee}, \citenamefont {Wagner}, \citenamefont {Bergholm},
  \citenamefont {Schulte-Herbr{\"u}ggen}, \citenamefont {Neumann},\ and\
  \citenamefont {Wrachtrup}}]{zaiser2016enhancing}%
  \BibitemOpen
  \bibfield  {author} {\bibinfo {author} {\bibfnamefont {S.}~\bibnamefont
  {Zaiser}}, \bibinfo {author} {\bibfnamefont {T.}~\bibnamefont {Rendler}},
  \bibinfo {author} {\bibfnamefont {I.}~\bibnamefont {Jakobi}}, \bibinfo
  {author} {\bibfnamefont {T.}~\bibnamefont {Wolf}}, \bibinfo {author}
  {\bibfnamefont {S.-Y.}\ \bibnamefont {Lee}}, \bibinfo {author} {\bibfnamefont
  {S.}~\bibnamefont {Wagner}}, \bibinfo {author} {\bibfnamefont
  {V.}~\bibnamefont {Bergholm}}, \bibinfo {author} {\bibfnamefont
  {T.}~\bibnamefont {Schulte-Herbr{\"u}ggen}}, \bibinfo {author} {\bibfnamefont
  {P.}~\bibnamefont {Neumann}},\ and\ \bibinfo {author} {\bibfnamefont
  {J.}~\bibnamefont {Wrachtrup}},\ }\bibfield  {title} {\bibinfo {title}
  {Enhancing quantum sensing sensitivity by a quantum memory},\ }\href@noop {}
  {\bibfield  {journal} {\bibinfo  {journal} {Nat. Commun.}\ }\textbf {\bibinfo
  {volume} {7}},\ \bibinfo {pages} {12279} (\bibinfo {year}
  {2016})}\BibitemShut {NoStop}%
\bibitem [{\citenamefont {Andrew}\ \emph {et~al.}(1958)\citenamefont {Andrew},
  \citenamefont {Bradbury},\ and\ \citenamefont {Eades}}]{andrew1958nuclear}%
  \BibitemOpen
  \bibfield  {author} {\bibinfo {author} {\bibfnamefont {E.}~\bibnamefont
  {Andrew}}, \bibinfo {author} {\bibfnamefont {A.}~\bibnamefont {Bradbury}},\
  and\ \bibinfo {author} {\bibfnamefont {R.}~\bibnamefont {Eades}},\ }\bibfield
   {title} {\bibinfo {title} {Nuclear magnetic resonance spectra from a crystal
  rotated at high speed},\ }\href@noop {} {\bibfield  {journal} {\bibinfo
  {journal} {Nature}\ }\textbf {\bibinfo {volume} {182}},\ \bibinfo {pages}
  {1659} (\bibinfo {year} {1958})}\BibitemShut {NoStop}%
\bibitem [{\citenamefont {Pykett}\ \emph {et~al.}(1982)\citenamefont {Pykett},
  \citenamefont {Newhouse}, \citenamefont {Buonanno}, \citenamefont {Brady},
  \citenamefont {Goldman}, \citenamefont {Kistler},\ and\ \citenamefont
  {Pohost}}]{pykett1982principles}%
  \BibitemOpen
  \bibfield  {author} {\bibinfo {author} {\bibfnamefont {I.~L.}\ \bibnamefont
  {Pykett}}, \bibinfo {author} {\bibfnamefont {J.~H.}\ \bibnamefont
  {Newhouse}}, \bibinfo {author} {\bibfnamefont {F.~S.}\ \bibnamefont
  {Buonanno}}, \bibinfo {author} {\bibfnamefont {T.~J.}\ \bibnamefont {Brady}},
  \bibinfo {author} {\bibfnamefont {M.~R.}\ \bibnamefont {Goldman}}, \bibinfo
  {author} {\bibfnamefont {J.~P.}\ \bibnamefont {Kistler}},\ and\ \bibinfo
  {author} {\bibfnamefont {G.~M.}\ \bibnamefont {Pohost}},\ }\bibfield  {title}
  {\bibinfo {title} {Principles of nuclear magnetic resonance imaging.},\
  }\href@noop {} {\bibfield  {journal} {\bibinfo  {journal} {Radiology}\
  }\textbf {\bibinfo {volume} {143}},\ \bibinfo {pages} {157} (\bibinfo {year}
  {1982})}\BibitemShut {NoStop}%
\bibitem [{\citenamefont {Lambert}\ \emph {et~al.}(2019)\citenamefont
  {Lambert}, \citenamefont {Mazzola},\ and\ \citenamefont
  {Ridge}}]{lambert2019nuclear}%
  \BibitemOpen
  \bibfield  {author} {\bibinfo {author} {\bibfnamefont {J.~B.}\ \bibnamefont
  {Lambert}}, \bibinfo {author} {\bibfnamefont {E.~P.}\ \bibnamefont
  {Mazzola}},\ and\ \bibinfo {author} {\bibfnamefont {C.~D.}\ \bibnamefont
  {Ridge}},\ }\href@noop {} {\emph {\bibinfo {title} {Nuclear magnetic
  resonance spectroscopy: an introduction to principles, applications, and
  experimental methods}}}\ (\bibinfo  {publisher} {John Wiley \& Sons},\
  \bibinfo {year} {2019})\BibitemShut {NoStop}%
\bibitem [{\citenamefont {Coronado}(2020)}]{coronado2020molecular}%
  \BibitemOpen
  \bibfield  {author} {\bibinfo {author} {\bibfnamefont {E.}~\bibnamefont
  {Coronado}},\ }\bibfield  {title} {\bibinfo {title} {Molecular magnetism:
  from chemical design to spin control in molecules, materials and devices},\
  }\href@noop {} {\bibfield  {journal} {\bibinfo  {journal} {Nat. Rev. Mater.}\
  }\textbf {\bibinfo {volume} {5}},\ \bibinfo {pages} {87} (\bibinfo {year}
  {2020})}\BibitemShut {NoStop}%
\bibitem [{\citenamefont {Gao}\ \emph {et~al.}(2022)\citenamefont {Gao},
  \citenamefont {Vaidya}, \citenamefont {Li}, \citenamefont {Ju}, \citenamefont
  {Jiang}, \citenamefont {Xu}, \citenamefont {Allcca}, \citenamefont {Shen},
  \citenamefont {Taniguchi}, \citenamefont {Watanabe} \emph
  {et~al.}}]{gao2022nuclear}%
  \BibitemOpen
  \bibfield  {author} {\bibinfo {author} {\bibfnamefont {X.}~\bibnamefont
  {Gao}}, \bibinfo {author} {\bibfnamefont {S.}~\bibnamefont {Vaidya}},
  \bibinfo {author} {\bibfnamefont {K.}~\bibnamefont {Li}}, \bibinfo {author}
  {\bibfnamefont {P.}~\bibnamefont {Ju}}, \bibinfo {author} {\bibfnamefont
  {B.}~\bibnamefont {Jiang}}, \bibinfo {author} {\bibfnamefont
  {Z.}~\bibnamefont {Xu}}, \bibinfo {author} {\bibfnamefont {A.~E.~L.}\
  \bibnamefont {Allcca}}, \bibinfo {author} {\bibfnamefont {K.}~\bibnamefont
  {Shen}}, \bibinfo {author} {\bibfnamefont {T.}~\bibnamefont {Taniguchi}},
  \bibinfo {author} {\bibfnamefont {K.}~\bibnamefont {Watanabe}}, \emph
  {et~al.},\ }\bibfield  {title} {\bibinfo {title} {Nuclear spin polarization
  and control in hexagonal boron nitride},\ }\href@noop {} {\bibfield
  {journal} {\bibinfo  {journal} {Nat. Mater.}\ }\textbf {\bibinfo {volume}
  {21}},\ \bibinfo {pages} {1024} (\bibinfo {year} {2022})}\BibitemShut
  {NoStop}%
\bibitem [{\citenamefont {Peik}\ and\ \citenamefont
  {Tamm}(2003)}]{peik2003nuclear}%
  \BibitemOpen
  \bibfield  {author} {\bibinfo {author} {\bibfnamefont {E.}~\bibnamefont
  {Peik}}\ and\ \bibinfo {author} {\bibfnamefont {C.}~\bibnamefont {Tamm}},\
  }\bibfield  {title} {\bibinfo {title} {Nuclear laser spectroscopy of the 3.5
  e{V} transition in {T}h-229},\ }\href@noop {} {\bibfield  {journal} {\bibinfo
   {journal} {Europhys. Lett.}\ }\textbf {\bibinfo {volume} {61}},\ \bibinfo
  {pages} {181} (\bibinfo {year} {2003})}\BibitemShut {NoStop}%
\bibitem [{\citenamefont {Tiedau}\ \emph {et~al.}(2024)\citenamefont {Tiedau},
  \citenamefont {Okhapkin}, \citenamefont {Zhang}, \citenamefont {Thielking},
  \citenamefont {Zitzer}, \citenamefont {Peik}, \citenamefont {Schaden},
  \citenamefont {Pronebner}, \citenamefont {Morawetz}, \citenamefont {De~Col}
  \emph {et~al.}}]{tiedau2024laser}%
  \BibitemOpen
  \bibfield  {author} {\bibinfo {author} {\bibfnamefont {J.}~\bibnamefont
  {Tiedau}}, \bibinfo {author} {\bibfnamefont {M.}~\bibnamefont {Okhapkin}},
  \bibinfo {author} {\bibfnamefont {K.}~\bibnamefont {Zhang}}, \bibinfo
  {author} {\bibfnamefont {J.}~\bibnamefont {Thielking}}, \bibinfo {author}
  {\bibfnamefont {G.}~\bibnamefont {Zitzer}}, \bibinfo {author} {\bibfnamefont
  {E.}~\bibnamefont {Peik}}, \bibinfo {author} {\bibfnamefont {F.}~\bibnamefont
  {Schaden}}, \bibinfo {author} {\bibfnamefont {T.}~\bibnamefont {Pronebner}},
  \bibinfo {author} {\bibfnamefont {I.}~\bibnamefont {Morawetz}}, \bibinfo
  {author} {\bibfnamefont {L.~T.}\ \bibnamefont {De~Col}}, \emph {et~al.},\
  }\bibfield  {title} {\bibinfo {title} {Laser excitation of the {T}h-229
  nucleus},\ }\href@noop {} {\bibfield  {journal} {\bibinfo  {journal} {Phys.
  Rev. Lett.}\ }\textbf {\bibinfo {volume} {132}},\ \bibinfo {pages} {182501}
  (\bibinfo {year} {2024})}\BibitemShut {NoStop}%
\bibitem [{\citenamefont {Walker}\ and\ \citenamefont
  {Dracoulis}(1999)}]{walker1999energy}%
  \BibitemOpen
  \bibfield  {author} {\bibinfo {author} {\bibfnamefont {P.}~\bibnamefont
  {Walker}}\ and\ \bibinfo {author} {\bibfnamefont {G.}~\bibnamefont
  {Dracoulis}},\ }\bibfield  {title} {\bibinfo {title} {Energy traps in atomic
  nuclei},\ }\href@noop {} {\bibfield  {journal} {\bibinfo  {journal} {Nature}\
  }\textbf {\bibinfo {volume} {399}},\ \bibinfo {pages} {35} (\bibinfo {year}
  {1999})}\BibitemShut {NoStop}%
\bibitem [{\citenamefont {Chiara}\ \emph {et~al.}(2018)\citenamefont {Chiara},
  \citenamefont {Carroll}, \citenamefont {Carpenter}, \citenamefont {Greene},
  \citenamefont {Hartley}, \citenamefont {Janssens}, \citenamefont {Lane},
  \citenamefont {Marsh}, \citenamefont {Matters}, \citenamefont {Polasik} \emph
  {et~al.}}]{chiara2018isomer}%
  \BibitemOpen
  \bibfield  {author} {\bibinfo {author} {\bibfnamefont {C.}~\bibnamefont
  {Chiara}}, \bibinfo {author} {\bibfnamefont {J.}~\bibnamefont {Carroll}},
  \bibinfo {author} {\bibfnamefont {M.}~\bibnamefont {Carpenter}}, \bibinfo
  {author} {\bibfnamefont {J.}~\bibnamefont {Greene}}, \bibinfo {author}
  {\bibfnamefont {D.}~\bibnamefont {Hartley}}, \bibinfo {author} {\bibfnamefont
  {R.}~\bibnamefont {Janssens}}, \bibinfo {author} {\bibfnamefont
  {G.}~\bibnamefont {Lane}}, \bibinfo {author} {\bibfnamefont {J.}~\bibnamefont
  {Marsh}}, \bibinfo {author} {\bibfnamefont {D.}~\bibnamefont {Matters}},
  \bibinfo {author} {\bibfnamefont {M.}~\bibnamefont {Polasik}}, \emph
  {et~al.},\ }\bibfield  {title} {\bibinfo {title} {Isomer depletion as
  experimental evidence of nuclear excitation by electron capture},\
  }\href@noop {} {\bibfield  {journal} {\bibinfo  {journal} {Nature}\ }\textbf
  {\bibinfo {volume} {554}},\ \bibinfo {pages} {216} (\bibinfo {year}
  {2018})}\BibitemShut {NoStop}%
\bibitem [{\citenamefont {Baldwin}\ and\ \citenamefont
  {Solem}(1997)}]{baldwin1997recoilless}%
  \BibitemOpen
  \bibfield  {author} {\bibinfo {author} {\bibfnamefont {G.~C.}\ \bibnamefont
  {Baldwin}}\ and\ \bibinfo {author} {\bibfnamefont {J.~C.}\ \bibnamefont
  {Solem}},\ }\bibfield  {title} {\bibinfo {title} {Recoilless gamma-ray
  lasers},\ }\href@noop {} {\bibfield  {journal} {\bibinfo  {journal} {Rev.
  Mod. Phys.}\ }\textbf {\bibinfo {volume} {69}},\ \bibinfo {pages} {1085}
  (\bibinfo {year} {1997})}\BibitemShut {NoStop}%
\bibitem [{\citenamefont {Tkalya}(2011)}]{tkalya2011proposal}%
  \BibitemOpen
  \bibfield  {author} {\bibinfo {author} {\bibfnamefont {E.}~\bibnamefont
  {Tkalya}},\ }\bibfield  {title} {\bibinfo {title} {Proposal for a nuclear
  gamma-ray laser of optical range},\ }\href@noop {} {\bibfield  {journal}
  {\bibinfo  {journal} {Phys. Rev. Lett.}\ }\textbf {\bibinfo {volume} {106}},\
  \bibinfo {pages} {162501} (\bibinfo {year} {2011})}\BibitemShut {NoStop}%
\bibitem [{\citenamefont {Zilges}\ \emph {et~al.}(2022)\citenamefont {Zilges},
  \citenamefont {Balabanski}, \citenamefont {Isaak},\ and\ \citenamefont
  {Pietralla}}]{zilges2022photonuclear}%
  \BibitemOpen
  \bibfield  {author} {\bibinfo {author} {\bibfnamefont {A.}~\bibnamefont
  {Zilges}}, \bibinfo {author} {\bibfnamefont {D.}~\bibnamefont {Balabanski}},
  \bibinfo {author} {\bibfnamefont {J.}~\bibnamefont {Isaak}},\ and\ \bibinfo
  {author} {\bibfnamefont {N.}~\bibnamefont {Pietralla}},\ }\bibfield  {title}
  {\bibinfo {title} {Photonuclear reactions—from basic research to
  applications},\ }\href@noop {} {\bibfield  {journal} {\bibinfo  {journal}
  {Prog. Part. Nucl. Phys.}\ }\textbf {\bibinfo {volume} {122}},\ \bibinfo
  {pages} {103903} (\bibinfo {year} {2022})}\BibitemShut {NoStop}%
\bibitem [{\citenamefont {Downer}\ \emph {et~al.}(2018)\citenamefont {Downer},
  \citenamefont {Zgadzaj}, \citenamefont {Debus}, \citenamefont {Schramm},\
  and\ \citenamefont {Kaluza}}]{downer2018diagnostics}%
  \BibitemOpen
  \bibfield  {author} {\bibinfo {author} {\bibfnamefont {M.~C.}\ \bibnamefont
  {Downer}}, \bibinfo {author} {\bibfnamefont {R.}~\bibnamefont {Zgadzaj}},
  \bibinfo {author} {\bibfnamefont {A.}~\bibnamefont {Debus}}, \bibinfo
  {author} {\bibfnamefont {U.}~\bibnamefont {Schramm}},\ and\ \bibinfo {author}
  {\bibfnamefont {M.}~\bibnamefont {Kaluza}},\ }\bibfield  {title} {\bibinfo
  {title} {Diagnostics for plasma-based electron accelerators},\ }\href@noop {}
  {\bibfield  {journal} {\bibinfo  {journal} {Rev. Mod. Phys.}\ }\textbf
  {\bibinfo {volume} {90}},\ \bibinfo {pages} {035002} (\bibinfo {year}
  {2018})}\BibitemShut {NoStop}%
\bibitem [{\citenamefont {Dubbers}\ and\ \citenamefont
  {Schmidt}(2011)}]{dubbers2011neutron}%
  \BibitemOpen
  \bibfield  {author} {\bibinfo {author} {\bibfnamefont {D.}~\bibnamefont
  {Dubbers}}\ and\ \bibinfo {author} {\bibfnamefont {M.~G.}\ \bibnamefont
  {Schmidt}},\ }\bibfield  {title} {\bibinfo {title} {The neutron and its role
  in cosmology and particle physics},\ }\href@noop {} {\bibfield  {journal}
  {\bibinfo  {journal} {Rev. Mod. Phys.}\ }\textbf {\bibinfo {volume} {83}},\
  \bibinfo {pages} {1111} (\bibinfo {year} {2011})}\BibitemShut {NoStop}%
\bibitem [{\citenamefont {Shiltsev}\ and\ \citenamefont
  {Zimmermann}(2021)}]{shiltsev2021modern}%
  \BibitemOpen
  \bibfield  {author} {\bibinfo {author} {\bibfnamefont {V.}~\bibnamefont
  {Shiltsev}}\ and\ \bibinfo {author} {\bibfnamefont {F.}~\bibnamefont
  {Zimmermann}},\ }\bibfield  {title} {\bibinfo {title} {Modern and future
  colliders},\ }\href@noop {} {\bibfield  {journal} {\bibinfo  {journal} {Rev.
  Mod. Phys.}\ }\textbf {\bibinfo {volume} {93}},\ \bibinfo {pages} {015006}
  (\bibinfo {year} {2021})}\BibitemShut {NoStop}%
\bibitem [{\citenamefont {Balabanski}\ \emph {et~al.}(2017)\citenamefont
  {Balabanski}, \citenamefont {Popescu}, \citenamefont {Stutman}, \citenamefont
  {Tanaka}, \citenamefont {Tesileanu}, \citenamefont {Ur}, \citenamefont
  {Ursescu},\ and\ \citenamefont {Zamfir}}]{balabanski2017new}%
  \BibitemOpen
  \bibfield  {author} {\bibinfo {author} {\bibfnamefont {D.}~\bibnamefont
  {Balabanski}}, \bibinfo {author} {\bibfnamefont {R.}~\bibnamefont {Popescu}},
  \bibinfo {author} {\bibfnamefont {D.}~\bibnamefont {Stutman}}, \bibinfo
  {author} {\bibfnamefont {K.}~\bibnamefont {Tanaka}}, \bibinfo {author}
  {\bibfnamefont {O.}~\bibnamefont {Tesileanu}}, \bibinfo {author}
  {\bibfnamefont {C.}~\bibnamefont {Ur}}, \bibinfo {author} {\bibfnamefont
  {D.}~\bibnamefont {Ursescu}},\ and\ \bibinfo {author} {\bibfnamefont
  {N.}~\bibnamefont {Zamfir}},\ }\bibfield  {title} {\bibinfo {title} {New
  light in nuclear physics: {T}he extreme light infrastructure},\ }\href@noop
  {} {\bibfield  {journal} {\bibinfo  {journal} {Europhys. Lett.}\ }\textbf
  {\bibinfo {volume} {117}},\ \bibinfo {pages} {28001} (\bibinfo {year}
  {2017})}\BibitemShut {NoStop}%
\bibitem [{\citenamefont {Zhang}\ \emph {et~al.}(2020)\citenamefont {Zhang},
  \citenamefont {Wu}, \citenamefont {Hu}, \citenamefont {Yang}, \citenamefont
  {Gui}, \citenamefont {Ji}, \citenamefont {Liu}, \citenamefont {Wang},
  \citenamefont {Liu}, \citenamefont {Lu} \emph {et~al.}}]{zhang2020laser}%
  \BibitemOpen
  \bibfield  {author} {\bibinfo {author} {\bibfnamefont {Z.}~\bibnamefont
  {Zhang}}, \bibinfo {author} {\bibfnamefont {F.}~\bibnamefont {Wu}}, \bibinfo
  {author} {\bibfnamefont {J.}~\bibnamefont {Hu}}, \bibinfo {author}
  {\bibfnamefont {X.}~\bibnamefont {Yang}}, \bibinfo {author} {\bibfnamefont
  {J.}~\bibnamefont {Gui}}, \bibinfo {author} {\bibfnamefont {P.}~\bibnamefont
  {Ji}}, \bibinfo {author} {\bibfnamefont {X.}~\bibnamefont {Liu}}, \bibinfo
  {author} {\bibfnamefont {C.}~\bibnamefont {Wang}}, \bibinfo {author}
  {\bibfnamefont {Y.}~\bibnamefont {Liu}}, \bibinfo {author} {\bibfnamefont
  {X.}~\bibnamefont {Lu}}, \emph {et~al.},\ }\bibfield  {title} {\bibinfo
  {title} {The laser beamline in {SULF} facility},\ }\href@noop {} {\bibfield
  {journal} {\bibinfo  {journal} {High Power Laser Sci. Eng.}\ }\textbf
  {\bibinfo {volume} {8}},\ \bibinfo {pages} {e4} (\bibinfo {year}
  {2020})}\BibitemShut {NoStop}%
\bibitem [{\citenamefont {Yoon}\ \emph {et~al.}(2021)\citenamefont {Yoon},
  \citenamefont {Kim}, \citenamefont {Choi}, \citenamefont {Sung},
  \citenamefont {Lee}, \citenamefont {Lee},\ and\ \citenamefont
  {Nam}}]{yoon2021realization}%
  \BibitemOpen
  \bibfield  {author} {\bibinfo {author} {\bibfnamefont {J.~W.}\ \bibnamefont
  {Yoon}}, \bibinfo {author} {\bibfnamefont {Y.~G.}\ \bibnamefont {Kim}},
  \bibinfo {author} {\bibfnamefont {I.~W.}\ \bibnamefont {Choi}}, \bibinfo
  {author} {\bibfnamefont {J.~H.}\ \bibnamefont {Sung}}, \bibinfo {author}
  {\bibfnamefont {H.~W.}\ \bibnamefont {Lee}}, \bibinfo {author} {\bibfnamefont
  {S.~K.}\ \bibnamefont {Lee}},\ and\ \bibinfo {author} {\bibfnamefont {C.~H.}\
  \bibnamefont {Nam}},\ }\bibfield  {title} {\bibinfo {title} {Realization of
  laser intensity over $10^{23}$ {W}/cm$^2$},\ }\href@noop {} {\bibfield
  {journal} {\bibinfo  {journal} {Optica}\ }\textbf {\bibinfo {volume} {8}},\
  \bibinfo {pages} {630} (\bibinfo {year} {2021})}\BibitemShut {NoStop}%
\bibitem [{\citenamefont {Feng}\ \emph {et~al.}(2022)\citenamefont {Feng},
  \citenamefont {Wang}, \citenamefont {Fu}, \citenamefont {Chen}, \citenamefont
  {Tan}, \citenamefont {Li}, \citenamefont {Wang}, \citenamefont {Li},
  \citenamefont {Zhang}, \citenamefont {Ma} \emph
  {et~al.}}]{feng2022femtosecond}%
  \BibitemOpen
  \bibfield  {author} {\bibinfo {author} {\bibfnamefont {J.}~\bibnamefont
  {Feng}}, \bibinfo {author} {\bibfnamefont {W.}~\bibnamefont {Wang}}, \bibinfo
  {author} {\bibfnamefont {C.}~\bibnamefont {Fu}}, \bibinfo {author}
  {\bibfnamefont {L.}~\bibnamefont {Chen}}, \bibinfo {author} {\bibfnamefont
  {J.}~\bibnamefont {Tan}}, \bibinfo {author} {\bibfnamefont {Y.}~\bibnamefont
  {Li}}, \bibinfo {author} {\bibfnamefont {J.}~\bibnamefont {Wang}}, \bibinfo
  {author} {\bibfnamefont {Y.}~\bibnamefont {Li}}, \bibinfo {author}
  {\bibfnamefont {G.}~\bibnamefont {Zhang}}, \bibinfo {author} {\bibfnamefont
  {Y.}~\bibnamefont {Ma}}, \emph {et~al.},\ }\bibfield  {title} {\bibinfo
  {title} {Femtosecond pumping of nuclear isomeric states by the {C}oulomb
  collision of ions with quivering electrons},\ }\href@noop {} {\bibfield
  {journal} {\bibinfo  {journal} {Phys. Rev. Lett.}\ }\textbf {\bibinfo
  {volume} {128}},\ \bibinfo {pages} {052501} (\bibinfo {year}
  {2022})}\BibitemShut {NoStop}%
\bibitem [{\citenamefont {Feng}\ \emph {et~al.}(2024)\citenamefont {Feng},
  \citenamefont {Qi}, \citenamefont {Zhang}, \citenamefont {Chen},
  \citenamefont {Zhu}, \citenamefont {Hu}, \citenamefont {Xu}, \citenamefont
  {Fu}, \citenamefont {Wang}, \citenamefont {Chen} \emph
  {et~al.}}]{feng2024laser}%
  \BibitemOpen
  \bibfield  {author} {\bibinfo {author} {\bibfnamefont {J.}~\bibnamefont
  {Feng}}, \bibinfo {author} {\bibfnamefont {J.}~\bibnamefont {Qi}}, \bibinfo
  {author} {\bibfnamefont {H.}~\bibnamefont {Zhang}}, \bibinfo {author}
  {\bibfnamefont {S.}~\bibnamefont {Chen}}, \bibinfo {author} {\bibfnamefont
  {M.}~\bibnamefont {Zhu}}, \bibinfo {author} {\bibfnamefont {X.}~\bibnamefont
  {Hu}}, \bibinfo {author} {\bibfnamefont {H.}~\bibnamefont {Xu}}, \bibinfo
  {author} {\bibfnamefont {C.}~\bibnamefont {Fu}}, \bibinfo {author}
  {\bibfnamefont {X.}~\bibnamefont {Wang}}, \bibinfo {author} {\bibfnamefont
  {L.}~\bibnamefont {Chen}}, \emph {et~al.},\ }\bibfield  {title} {\bibinfo
  {title} {Laser-based approach to measure small nuclear cross sections in
  plasma},\ }\href@noop {} {\bibfield  {journal} {\bibinfo  {journal} {P. Natl.
  Acad. Sci. USA}\ }\textbf {\bibinfo {volume} {121}},\ \bibinfo {pages}
  {e2413221121} (\bibinfo {year} {2024})}\BibitemShut {NoStop}%
\bibitem [{\citenamefont {Gargiulo}\ \emph {et~al.}(2024)\citenamefont
  {Gargiulo}, \citenamefont {Madan}, \citenamefont {Truc}, \citenamefont
  {Usai}, \citenamefont {Beeks}, \citenamefont {Leccese}, \citenamefont
  {Vanacore},\ and\ \citenamefont {Carbone}}]{gargiulo2024revisiting}%
  \BibitemOpen
  \bibfield  {author} {\bibinfo {author} {\bibfnamefont {S.}~\bibnamefont
  {Gargiulo}}, \bibinfo {author} {\bibfnamefont {I.}~\bibnamefont {Madan}},
  \bibinfo {author} {\bibfnamefont {B.}~\bibnamefont {Truc}}, \bibinfo {author}
  {\bibfnamefont {P.}~\bibnamefont {Usai}}, \bibinfo {author} {\bibfnamefont
  {K.}~\bibnamefont {Beeks}}, \bibinfo {author} {\bibfnamefont
  {V.}~\bibnamefont {Leccese}}, \bibinfo {author} {\bibfnamefont {G.~M.}\
  \bibnamefont {Vanacore}},\ and\ \bibinfo {author} {\bibfnamefont
  {F.}~\bibnamefont {Carbone}},\ }\bibfield  {title} {\bibinfo {title}
  {Revisiting the {E}xcitation of the {L}ow-{L}ying $^{181m}${T}a {I}somer in
  {O}ptical {L}aser-{G}enerated {P}lasma},\ }\href@noop {} {\bibfield
  {journal} {\bibinfo  {journal} {Phys. Rev. Lett.}\ }\textbf {\bibinfo
  {volume} {133}},\ \bibinfo {pages} {132501} (\bibinfo {year}
  {2024})}\BibitemShut {NoStop}%
\bibitem [{\citenamefont {Jacob}\ \emph {et~al.}(2025)\citenamefont {Jacob},
  \citenamefont {Tannous}, \citenamefont {Bernstein}, \citenamefont {Brown},
  \citenamefont {Ostermayr}, \citenamefont {Chen}, \citenamefont {Schneider},
  \citenamefont {Schroeder}, \citenamefont {van Tilborg}, \citenamefont
  {Esarey} \emph {et~al.}}]{jacob2025enhanced}%
  \BibitemOpen
  \bibfield  {author} {\bibinfo {author} {\bibfnamefont {R.~E.}\ \bibnamefont
  {Jacob}}, \bibinfo {author} {\bibfnamefont {S.~M.}\ \bibnamefont {Tannous}},
  \bibinfo {author} {\bibfnamefont {L.~A.}\ \bibnamefont {Bernstein}}, \bibinfo
  {author} {\bibfnamefont {J.}~\bibnamefont {Brown}}, \bibinfo {author}
  {\bibfnamefont {T.}~\bibnamefont {Ostermayr}}, \bibinfo {author}
  {\bibfnamefont {Q.}~\bibnamefont {Chen}}, \bibinfo {author} {\bibfnamefont
  {D.~H.~G.}\ \bibnamefont {Schneider}}, \bibinfo {author} {\bibfnamefont
  {C.~B.}\ \bibnamefont {Schroeder}}, \bibinfo {author} {\bibfnamefont
  {J.}~\bibnamefont {van Tilborg}}, \bibinfo {author} {\bibfnamefont {E.~H.}\
  \bibnamefont {Esarey}}, \emph {et~al.},\ }\bibfield  {title} {\bibinfo
  {title} {Enhanced {I}somer {P}opulation via {D}irect {I}rradiation of
  {S}olid-{D}ensity {T}argets {U}sing a {C}ompact {L}aser-{P}lasma
  {A}ccelerator},\ }\href@noop {} {\bibfield  {journal} {\bibinfo  {journal}
  {Phys. Rev. Lett.}\ }\textbf {\bibinfo {volume} {134}},\ \bibinfo {pages}
  {052504} (\bibinfo {year} {2025})}\BibitemShut {NoStop}%
\bibitem [{\citenamefont {Wang}\ \emph {et~al.}(2021)\citenamefont {Wang},
  \citenamefont {Zhou}, \citenamefont {Liu},\ and\ \citenamefont
  {Wang}}]{wang2021exciting}%
  \BibitemOpen
  \bibfield  {author} {\bibinfo {author} {\bibfnamefont {W.}~\bibnamefont
  {Wang}}, \bibinfo {author} {\bibfnamefont {J.}~\bibnamefont {Zhou}}, \bibinfo
  {author} {\bibfnamefont {B.}~\bibnamefont {Liu}},\ and\ \bibinfo {author}
  {\bibfnamefont {X.}~\bibnamefont {Wang}},\ }\bibfield  {title} {\bibinfo
  {title} {Exciting the isomeric $^{229}${T}h nuclear state via laser-driven
  electron recollision},\ }\href@noop {} {\bibfield  {journal} {\bibinfo
  {journal} {Phys. Rev. Lett.}\ }\textbf {\bibinfo {volume} {127}},\ \bibinfo
  {pages} {052501} (\bibinfo {year} {2021})}\BibitemShut {NoStop}%
\bibitem [{\citenamefont {Qi}\ \emph {et~al.}(2023)\citenamefont {Qi},
  \citenamefont {Zhang},\ and\ \citenamefont {Wang}}]{qi2023isomeric}%
  \BibitemOpen
  \bibfield  {author} {\bibinfo {author} {\bibfnamefont {J.}~\bibnamefont
  {Qi}}, \bibinfo {author} {\bibfnamefont {H.}~\bibnamefont {Zhang}},\ and\
  \bibinfo {author} {\bibfnamefont {X.}~\bibnamefont {Wang}},\ }\bibfield
  {title} {\bibinfo {title} {Isomeric excitation of $^{229}${T}h in
  laser-heated clusters},\ }\href@noop {} {\bibfield  {journal} {\bibinfo
  {journal} {Phys. Rev. Lett.}\ }\textbf {\bibinfo {volume} {130}},\ \bibinfo
  {pages} {112501} (\bibinfo {year} {2023})}\BibitemShut {NoStop}%
\bibitem [{\citenamefont {Wang}\ \emph {et~al.}(2024)\citenamefont {Wang},
  \citenamefont {Zou}, \citenamefont {Fritzsche},\ and\ \citenamefont
  {Li}}]{wang2024isomeric}%
  \BibitemOpen
  \bibfield  {author} {\bibinfo {author} {\bibfnamefont {W.}~\bibnamefont
  {Wang}}, \bibinfo {author} {\bibfnamefont {F.}~\bibnamefont {Zou}}, \bibinfo
  {author} {\bibfnamefont {S.}~\bibnamefont {Fritzsche}},\ and\ \bibinfo
  {author} {\bibfnamefont {Y.}~\bibnamefont {Li}},\ }\bibfield  {title}
  {\bibinfo {title} {Isomeric {P}opulation {T}ransfer of the $^{229}${T}h
  {N}ucleus via {H}yperfine {E}lectronic {B}ridge},\ }\href@noop {} {\bibfield
  {journal} {\bibinfo  {journal} {Phys. Rev. Lett.}\ }\textbf {\bibinfo
  {volume} {133}},\ \bibinfo {pages} {223001} (\bibinfo {year}
  {2024})}\BibitemShut {NoStop}%
\bibitem [{\citenamefont {Zhang}\ \emph {et~al.}(2024)\citenamefont {Zhang},
  \citenamefont {Li},\ and\ \citenamefont {Wang}}]{zhang2024highly}%
  \BibitemOpen
  \bibfield  {author} {\bibinfo {author} {\bibfnamefont {H.}~\bibnamefont
  {Zhang}}, \bibinfo {author} {\bibfnamefont {T.}~\bibnamefont {Li}},\ and\
  \bibinfo {author} {\bibfnamefont {X.}~\bibnamefont {Wang}},\ }\bibfield
  {title} {\bibinfo {title} {Highly nonlinear light-nucleus interaction},\
  }\href@noop {} {\bibfield  {journal} {\bibinfo  {journal} {Phys. Rev. Lett.}\
  }\textbf {\bibinfo {volume} {133}},\ \bibinfo {pages} {152503} (\bibinfo
  {year} {2024})}\BibitemShut {NoStop}%
\bibitem [{\citenamefont {Wycech}\ and\ \citenamefont
  {Zylicz}(1993)}]{wycech1993predictions}%
  \BibitemOpen
  \bibfield  {author} {\bibinfo {author} {\bibfnamefont {S.}~\bibnamefont
  {Wycech}}\ and\ \bibinfo {author} {\bibfnamefont {J.}~\bibnamefont
  {Zylicz}},\ }\bibfield  {title} {\bibinfo {title} {Predictions for nuclear
  spin mixing in magnetic fields},\ }\href@noop {} {\bibfield  {journal}
  {\bibinfo  {journal} {Acta Phys. Pol. B}\ }\textbf {\bibinfo {volume} {24}},\
  \bibinfo {pages} {637} (\bibinfo {year} {1993})}\BibitemShut {NoStop}%
\bibitem [{\citenamefont {Karpeshin}\ \emph {et~al.}(1998)\citenamefont
  {Karpeshin}, \citenamefont {Wycech}, \citenamefont {Band}, \citenamefont
  {Trzhaskovskaya}, \citenamefont {Pf{\"u}tzner},\ and\ \citenamefont
  {{\.Z}ylicz}}]{karpeshin1998rates}%
  \BibitemOpen
  \bibfield  {author} {\bibinfo {author} {\bibfnamefont {F.}~\bibnamefont
  {Karpeshin}}, \bibinfo {author} {\bibfnamefont {S.}~\bibnamefont {Wycech}},
  \bibinfo {author} {\bibfnamefont {I.}~\bibnamefont {Band}}, \bibinfo {author}
  {\bibfnamefont {M.}~\bibnamefont {Trzhaskovskaya}}, \bibinfo {author}
  {\bibfnamefont {M.}~\bibnamefont {Pf{\"u}tzner}},\ and\ \bibinfo {author}
  {\bibfnamefont {J.}~\bibnamefont {{\.Z}ylicz}},\ }\bibfield  {title}
  {\bibinfo {title} {Rates of transitions between the hyperfine-splitting
  components of the ground-state and the 3.5 e{V} isomer in
  $^{229}${T}h$^{89+}$},\ }\href@noop {} {\bibfield  {journal} {\bibinfo
  {journal} {Phys. Rev. C}\ }\textbf {\bibinfo {volume} {57}},\ \bibinfo
  {pages} {3085} (\bibinfo {year} {1998})}\BibitemShut {NoStop}%
\bibitem [{\citenamefont {Shabaev}\ \emph {et~al.}(2022)\citenamefont
  {Shabaev}, \citenamefont {Glazov}, \citenamefont {Ryzhkov}, \citenamefont
  {Brandau}, \citenamefont {Plunien}, \citenamefont {Quint}, \citenamefont
  {Volchkova},\ and\ \citenamefont {Zinenko}}]{shabaev2022ground}%
  \BibitemOpen
  \bibfield  {author} {\bibinfo {author} {\bibfnamefont {V.}~\bibnamefont
  {Shabaev}}, \bibinfo {author} {\bibfnamefont {D.}~\bibnamefont {Glazov}},
  \bibinfo {author} {\bibfnamefont {A.}~\bibnamefont {Ryzhkov}}, \bibinfo
  {author} {\bibfnamefont {C.}~\bibnamefont {Brandau}}, \bibinfo {author}
  {\bibfnamefont {G.}~\bibnamefont {Plunien}}, \bibinfo {author} {\bibfnamefont
  {W.}~\bibnamefont {Quint}}, \bibinfo {author} {\bibfnamefont
  {A.}~\bibnamefont {Volchkova}},\ and\ \bibinfo {author} {\bibfnamefont
  {D.}~\bibnamefont {Zinenko}},\ }\bibfield  {title} {\bibinfo {title}
  {{G}round-{S}tate $g$ {F}actor of {H}ighly {C}harged $^{229}${T}h {I}ons:
  {A}n {A}ccess to the {M}1 {T}ransition {P}robability between the {I}someric
  and {G}round {N}uclear {S}tates},\ }\href@noop {} {\bibfield  {journal}
  {\bibinfo  {journal} {Phys. Rev. Lett.}\ }\textbf {\bibinfo {volume} {128}},\
  \bibinfo {pages} {043001} (\bibinfo {year} {2022})}\BibitemShut {NoStop}%
\bibitem [{\citenamefont {Wang}\ and\ \citenamefont
  {Wang}(2024)}]{wang2024substantial}%
  \BibitemOpen
  \bibfield  {author} {\bibinfo {author} {\bibfnamefont {W.}~\bibnamefont
  {Wang}}\ and\ \bibinfo {author} {\bibfnamefont {X.}~\bibnamefont {Wang}},\
  }\bibfield  {title} {\bibinfo {title} {Substantial {N}uclear {H}yperfine
  {M}ixing effect in {B}oronlike $^{205}${P}b {I}ons},\ }\href@noop {}
  {\bibfield  {journal} {\bibinfo  {journal} {Phys. Rev. Lett.}\ }\textbf
  {\bibinfo {volume} {133}},\ \bibinfo {pages} {032501} (\bibinfo {year}
  {2024})}\BibitemShut {NoStop}%
\bibitem [{\citenamefont {Lee}\ \emph {et~al.}(2019)\citenamefont {Lee},
  \citenamefont {Alexander}, \citenamefont {Kevan}, \citenamefont {Roy},\ and\
  \citenamefont {McMorran}}]{lee2019laguerre}%
  \BibitemOpen
  \bibfield  {author} {\bibinfo {author} {\bibfnamefont {J.~T.}\ \bibnamefont
  {Lee}}, \bibinfo {author} {\bibfnamefont {S.}~\bibnamefont {Alexander}},
  \bibinfo {author} {\bibfnamefont {S.}~\bibnamefont {Kevan}}, \bibinfo
  {author} {\bibfnamefont {S.}~\bibnamefont {Roy}},\ and\ \bibinfo {author}
  {\bibfnamefont {B.}~\bibnamefont {McMorran}},\ }\bibfield  {title} {\bibinfo
  {title} {Laguerre--{G}auss and {H}ermite--{G}auss soft {X}-ray states
  generated using diffractive optics},\ }\href@noop {} {\bibfield  {journal}
  {\bibinfo  {journal} {Nat. Photonics}\ }\textbf {\bibinfo {volume} {13}},\
  \bibinfo {pages} {205} (\bibinfo {year} {2019})}\BibitemShut {NoStop}%
\bibitem [{\citenamefont {Luski}\ \emph {et~al.}(2021)\citenamefont {Luski},
  \citenamefont {Segev}, \citenamefont {David}, \citenamefont {Bitton},
  \citenamefont {Nadler}, \citenamefont {Barnea}, \citenamefont {Gorlach},
  \citenamefont {Cheshnovsky}, \citenamefont {Kaminer},\ and\ \citenamefont
  {Narevicius}}]{luski2021vortex}%
  \BibitemOpen
  \bibfield  {author} {\bibinfo {author} {\bibfnamefont {A.}~\bibnamefont
  {Luski}}, \bibinfo {author} {\bibfnamefont {Y.}~\bibnamefont {Segev}},
  \bibinfo {author} {\bibfnamefont {R.}~\bibnamefont {David}}, \bibinfo
  {author} {\bibfnamefont {O.}~\bibnamefont {Bitton}}, \bibinfo {author}
  {\bibfnamefont {H.}~\bibnamefont {Nadler}}, \bibinfo {author} {\bibfnamefont
  {A.~R.}\ \bibnamefont {Barnea}}, \bibinfo {author} {\bibfnamefont
  {A.}~\bibnamefont {Gorlach}}, \bibinfo {author} {\bibfnamefont
  {O.}~\bibnamefont {Cheshnovsky}}, \bibinfo {author} {\bibfnamefont
  {I.}~\bibnamefont {Kaminer}},\ and\ \bibinfo {author} {\bibfnamefont
  {E.}~\bibnamefont {Narevicius}},\ }\bibfield  {title} {\bibinfo {title}
  {Vortex beams of atoms and molecules},\ }\href@noop {} {\bibfield  {journal}
  {\bibinfo  {journal} {Science}\ }\textbf {\bibinfo {volume} {373}},\ \bibinfo
  {pages} {1105} (\bibinfo {year} {2021})}\BibitemShut {NoStop}%
\bibitem [{\citenamefont {Clark}\ \emph {et~al.}(2015)\citenamefont {Clark},
  \citenamefont {Barankov}, \citenamefont {Huber}, \citenamefont {Arif},
  \citenamefont {Cory},\ and\ \citenamefont {Pushin}}]{clark2015controlling}%
  \BibitemOpen
  \bibfield  {author} {\bibinfo {author} {\bibfnamefont {C.~W.}\ \bibnamefont
  {Clark}}, \bibinfo {author} {\bibfnamefont {R.}~\bibnamefont {Barankov}},
  \bibinfo {author} {\bibfnamefont {M.~G.}\ \bibnamefont {Huber}}, \bibinfo
  {author} {\bibfnamefont {M.}~\bibnamefont {Arif}}, \bibinfo {author}
  {\bibfnamefont {D.~G.}\ \bibnamefont {Cory}},\ and\ \bibinfo {author}
  {\bibfnamefont {D.~A.}\ \bibnamefont {Pushin}},\ }\bibfield  {title}
  {\bibinfo {title} {Controlling neutron orbital angular momentum},\
  }\href@noop {} {\bibfield  {journal} {\bibinfo  {journal} {Nature}\ }\textbf
  {\bibinfo {volume} {525}},\ \bibinfo {pages} {504} (\bibinfo {year}
  {2015})}\BibitemShut {NoStop}%
\bibitem [{\citenamefont {McMorran}\ \emph {et~al.}(2011)\citenamefont
  {McMorran}, \citenamefont {Agrawal}, \citenamefont {Anderson}, \citenamefont
  {Herzing}, \citenamefont {Lezec}, \citenamefont {McClelland},\ and\
  \citenamefont {Unguris}}]{mcmorran2011electron}%
  \BibitemOpen
  \bibfield  {author} {\bibinfo {author} {\bibfnamefont {B.~J.}\ \bibnamefont
  {McMorran}}, \bibinfo {author} {\bibfnamefont {A.}~\bibnamefont {Agrawal}},
  \bibinfo {author} {\bibfnamefont {I.~M.}\ \bibnamefont {Anderson}}, \bibinfo
  {author} {\bibfnamefont {A.~A.}\ \bibnamefont {Herzing}}, \bibinfo {author}
  {\bibfnamefont {H.~J.}\ \bibnamefont {Lezec}}, \bibinfo {author}
  {\bibfnamefont {J.~J.}\ \bibnamefont {McClelland}},\ and\ \bibinfo {author}
  {\bibfnamefont {J.}~\bibnamefont {Unguris}},\ }\bibfield  {title} {\bibinfo
  {title} {Electron vortex beams with high quanta of orbital angular
  momentum},\ }\href@noop {} {\bibfield  {journal} {\bibinfo  {journal}
  {Science}\ }\textbf {\bibinfo {volume} {331}},\ \bibinfo {pages} {192}
  (\bibinfo {year} {2011})}\BibitemShut {NoStop}%
\bibitem [{\citenamefont {Allen}\ \emph {et~al.}(1992)\citenamefont {Allen},
  \citenamefont {Beijersbergen}, \citenamefont {Spreeuw},\ and\ \citenamefont
  {Woerdman}}]{allen1992orbital}%
  \BibitemOpen
  \bibfield  {author} {\bibinfo {author} {\bibfnamefont {L.}~\bibnamefont
  {Allen}}, \bibinfo {author} {\bibfnamefont {M.~W.}\ \bibnamefont
  {Beijersbergen}}, \bibinfo {author} {\bibfnamefont {R.}~\bibnamefont
  {Spreeuw}},\ and\ \bibinfo {author} {\bibfnamefont {J.}~\bibnamefont
  {Woerdman}},\ }\bibfield  {title} {\bibinfo {title} {Orbital angular momentum
  of light and the transformation of {L}aguerre-{G}aussian laser modes},\
  }\href@noop {} {\bibfield  {journal} {\bibinfo  {journal} {Phys. Rev. A}\
  }\textbf {\bibinfo {volume} {45}},\ \bibinfo {pages} {8185} (\bibinfo {year}
  {1992})}\BibitemShut {NoStop}%
\bibitem [{\citenamefont {Knyazev}\ and\ \citenamefont
  {Serbo}(2018)}]{knyazev2018beams}%
  \BibitemOpen
  \bibfield  {author} {\bibinfo {author} {\bibfnamefont {B.~A.}\ \bibnamefont
  {Knyazev}}\ and\ \bibinfo {author} {\bibfnamefont {V.}~\bibnamefont
  {Serbo}},\ }\bibfield  {title} {\bibinfo {title} {Beams of photons with
  nonzero projections of orbital angular momenta: new results},\ }\href@noop {}
  {\bibfield  {journal} {\bibinfo  {journal} {Phys. Usp.}\ }\textbf {\bibinfo
  {volume} {61}},\ \bibinfo {pages} {449} (\bibinfo {year} {2018})}\BibitemShut
  {NoStop}%
\bibitem [{\citenamefont {Shen}\ \emph {et~al.}(2019)\citenamefont {Shen},
  \citenamefont {Wang}, \citenamefont {Xie}, \citenamefont {Min}, \citenamefont
  {Fu}, \citenamefont {Liu}, \citenamefont {Gong},\ and\ \citenamefont
  {Yuan}}]{shen2019optical}%
  \BibitemOpen
  \bibfield  {author} {\bibinfo {author} {\bibfnamefont {Y.}~\bibnamefont
  {Shen}}, \bibinfo {author} {\bibfnamefont {X.}~\bibnamefont {Wang}}, \bibinfo
  {author} {\bibfnamefont {Z.}~\bibnamefont {Xie}}, \bibinfo {author}
  {\bibfnamefont {C.}~\bibnamefont {Min}}, \bibinfo {author} {\bibfnamefont
  {X.}~\bibnamefont {Fu}}, \bibinfo {author} {\bibfnamefont {Q.}~\bibnamefont
  {Liu}}, \bibinfo {author} {\bibfnamefont {M.}~\bibnamefont {Gong}},\ and\
  \bibinfo {author} {\bibfnamefont {X.}~\bibnamefont {Yuan}},\ }\bibfield
  {title} {\bibinfo {title} {Optical vortices 30 years on: Oam manipulation
  from topological charge to multiple singularities},\ }\href@noop {}
  {\bibfield  {journal} {\bibinfo  {journal} {Light Sci. Appl.}\ }\textbf
  {\bibinfo {volume} {8}},\ \bibinfo {pages} {90} (\bibinfo {year}
  {2019})}\BibitemShut {NoStop}%
\bibitem [{\citenamefont {Peele}\ \emph {et~al.}(2002)\citenamefont {Peele},
  \citenamefont {McMahon}, \citenamefont {Paterson}, \citenamefont {Tran},
  \citenamefont {Mancuso}, \citenamefont {Nugent}, \citenamefont {Hayes},
  \citenamefont {Harvey}, \citenamefont {Lai},\ and\ \citenamefont
  {McNulty}}]{peele2002observation}%
  \BibitemOpen
  \bibfield  {author} {\bibinfo {author} {\bibfnamefont {A.~G.}\ \bibnamefont
  {Peele}}, \bibinfo {author} {\bibfnamefont {P.~J.}\ \bibnamefont {McMahon}},
  \bibinfo {author} {\bibfnamefont {D.}~\bibnamefont {Paterson}}, \bibinfo
  {author} {\bibfnamefont {C.~Q.}\ \bibnamefont {Tran}}, \bibinfo {author}
  {\bibfnamefont {A.~P.}\ \bibnamefont {Mancuso}}, \bibinfo {author}
  {\bibfnamefont {K.~A.}\ \bibnamefont {Nugent}}, \bibinfo {author}
  {\bibfnamefont {J.~P.}\ \bibnamefont {Hayes}}, \bibinfo {author}
  {\bibfnamefont {E.}~\bibnamefont {Harvey}}, \bibinfo {author} {\bibfnamefont
  {B.}~\bibnamefont {Lai}},\ and\ \bibinfo {author} {\bibfnamefont
  {I.}~\bibnamefont {McNulty}},\ }\bibfield  {title} {\bibinfo {title}
  {Observation of an x-ray vortex},\ }\href@noop {} {\bibfield  {journal}
  {\bibinfo  {journal} {Opt. Lett.}\ }\textbf {\bibinfo {volume} {27}},\
  \bibinfo {pages} {1752} (\bibinfo {year} {2002})}\BibitemShut {NoStop}%
\bibitem [{\citenamefont {Terhalle}\ \emph {et~al.}(2011)\citenamefont
  {Terhalle}, \citenamefont {Langner}, \citenamefont {P{\"a}iv{\"a}nranta},
  \citenamefont {Guzenko}, \citenamefont {David},\ and\ \citenamefont
  {Ekinci}}]{terhalle2011generation}%
  \BibitemOpen
  \bibfield  {author} {\bibinfo {author} {\bibfnamefont {B.}~\bibnamefont
  {Terhalle}}, \bibinfo {author} {\bibfnamefont {A.}~\bibnamefont {Langner}},
  \bibinfo {author} {\bibfnamefont {B.}~\bibnamefont {P{\"a}iv{\"a}nranta}},
  \bibinfo {author} {\bibfnamefont {V.~A.}\ \bibnamefont {Guzenko}}, \bibinfo
  {author} {\bibfnamefont {C.}~\bibnamefont {David}},\ and\ \bibinfo {author}
  {\bibfnamefont {Y.}~\bibnamefont {Ekinci}},\ }\bibfield  {title} {\bibinfo
  {title} {Generation of extreme ultraviolet vortex beams using computer
  generated holograms},\ }\href@noop {} {\bibfield  {journal} {\bibinfo
  {journal} {Opt. Lett.}\ }\textbf {\bibinfo {volume} {36}},\ \bibinfo {pages}
  {4143} (\bibinfo {year} {2011})}\BibitemShut {NoStop}%
\bibitem [{\citenamefont {Gariepy}\ \emph {et~al.}(2014)\citenamefont
  {Gariepy}, \citenamefont {Leach}, \citenamefont {Kim}, \citenamefont
  {Hammond}, \citenamefont {Frumker}, \citenamefont {Boyd},\ and\ \citenamefont
  {Corkum}}]{gariepy2014creating}%
  \BibitemOpen
  \bibfield  {author} {\bibinfo {author} {\bibfnamefont {G.}~\bibnamefont
  {Gariepy}}, \bibinfo {author} {\bibfnamefont {J.}~\bibnamefont {Leach}},
  \bibinfo {author} {\bibfnamefont {K.~T.}\ \bibnamefont {Kim}}, \bibinfo
  {author} {\bibfnamefont {T.~J.}\ \bibnamefont {Hammond}}, \bibinfo {author}
  {\bibfnamefont {E.}~\bibnamefont {Frumker}}, \bibinfo {author} {\bibfnamefont
  {R.~W.}\ \bibnamefont {Boyd}},\ and\ \bibinfo {author} {\bibfnamefont
  {P.~B.}\ \bibnamefont {Corkum}},\ }\bibfield  {title} {\bibinfo {title}
  {Creating high-harmonic beams with controlled orbital angular momentum},\
  }\href@noop {} {\bibfield  {journal} {\bibinfo  {journal} {Phys. Rev. Lett.}\
  }\textbf {\bibinfo {volume} {113}},\ \bibinfo {pages} {153901} (\bibinfo
  {year} {2014})}\BibitemShut {NoStop}%
\bibitem [{\citenamefont {Hemsing}\ \emph {et~al.}(2013)\citenamefont
  {Hemsing}, \citenamefont {Knyazik}, \citenamefont {Dunning}, \citenamefont
  {Xiang}, \citenamefont {Marinelli}, \citenamefont {Hast},\ and\ \citenamefont
  {Rosenzweig}}]{hemsing2013coherent}%
  \BibitemOpen
  \bibfield  {author} {\bibinfo {author} {\bibfnamefont {E.}~\bibnamefont
  {Hemsing}}, \bibinfo {author} {\bibfnamefont {A.}~\bibnamefont {Knyazik}},
  \bibinfo {author} {\bibfnamefont {M.}~\bibnamefont {Dunning}}, \bibinfo
  {author} {\bibfnamefont {D.}~\bibnamefont {Xiang}}, \bibinfo {author}
  {\bibfnamefont {A.}~\bibnamefont {Marinelli}}, \bibinfo {author}
  {\bibfnamefont {C.}~\bibnamefont {Hast}},\ and\ \bibinfo {author}
  {\bibfnamefont {J.~B.}\ \bibnamefont {Rosenzweig}},\ }\bibfield  {title}
  {\bibinfo {title} {Coherent optical vortices from relativistic electron
  beams},\ }\href@noop {} {\bibfield  {journal} {\bibinfo  {journal} {Nat.
  Phys.}\ }\textbf {\bibinfo {volume} {9}},\ \bibinfo {pages} {549} (\bibinfo
  {year} {2013})}\BibitemShut {NoStop}%
\bibitem [{\citenamefont {Ivanov}(2022)}]{ivanov2022promises}%
  \BibitemOpen
  \bibfield  {author} {\bibinfo {author} {\bibfnamefont {I.~P.}\ \bibnamefont
  {Ivanov}},\ }\bibfield  {title} {\bibinfo {title} {Promises and challenges of
  high-energy vortex states collisions},\ }\href@noop {} {\bibfield  {journal}
  {\bibinfo  {journal} {Prog. Part. Nucl. Phys.}\ }\textbf {\bibinfo {volume}
  {127}},\ \bibinfo {pages} {103987} (\bibinfo {year} {2022})}\BibitemShut
  {NoStop}%
\bibitem [{\citenamefont {Lange}\ \emph {et~al.}(2022)\citenamefont {Lange},
  \citenamefont {Huntemann}, \citenamefont {Peshkov}, \citenamefont
  {Surzhykov},\ and\ \citenamefont {Peik}}]{lange2022excitation}%
  \BibitemOpen
  \bibfield  {author} {\bibinfo {author} {\bibfnamefont {R.}~\bibnamefont
  {Lange}}, \bibinfo {author} {\bibfnamefont {N.}~\bibnamefont {Huntemann}},
  \bibinfo {author} {\bibfnamefont {A.}~\bibnamefont {Peshkov}}, \bibinfo
  {author} {\bibfnamefont {A.}~\bibnamefont {Surzhykov}},\ and\ \bibinfo
  {author} {\bibfnamefont {E.}~\bibnamefont {Peik}},\ }\bibfield  {title}
  {\bibinfo {title} {Excitation of an electric octupole transition by twisted
  light},\ }\href@noop {} {\bibfield  {journal} {\bibinfo  {journal} {Phys.
  Rev. Lett.}\ }\textbf {\bibinfo {volume} {129}},\ \bibinfo {pages} {253901}
  (\bibinfo {year} {2022})}\BibitemShut {NoStop}%
\bibitem [{\citenamefont {Afanasev}\ \emph {et~al.}(2018)\citenamefont
  {Afanasev}, \citenamefont {Carlson}, \citenamefont {Schmiegelow},
  \citenamefont {Schulz}, \citenamefont {Schmidt-Kaler},\ and\ \citenamefont
  {Solyanik}}]{afanasev2018experimental}%
  \BibitemOpen
  \bibfield  {author} {\bibinfo {author} {\bibfnamefont {A.}~\bibnamefont
  {Afanasev}}, \bibinfo {author} {\bibfnamefont {C.~E.}\ \bibnamefont
  {Carlson}}, \bibinfo {author} {\bibfnamefont {C.~T.}\ \bibnamefont
  {Schmiegelow}}, \bibinfo {author} {\bibfnamefont {J.}~\bibnamefont {Schulz}},
  \bibinfo {author} {\bibfnamefont {F.}~\bibnamefont {Schmidt-Kaler}},\ and\
  \bibinfo {author} {\bibfnamefont {M.}~\bibnamefont {Solyanik}},\ }\bibfield
  {title} {\bibinfo {title} {Experimental verification of position-dependent
  angular-momentum selection rules for absorption of twisted light by a bound
  electron},\ }\href@noop {} {\bibfield  {journal} {\bibinfo  {journal} {New J.
  Phys.}\ }\textbf {\bibinfo {volume} {20}},\ \bibinfo {pages} {023032}
  (\bibinfo {year} {2018})}\BibitemShut {NoStop}%
\bibitem [{\citenamefont {Schmiegelow}\ \emph {et~al.}(2016)\citenamefont
  {Schmiegelow}, \citenamefont {Schulz}, \citenamefont {Kaufmann},
  \citenamefont {Ruster}, \citenamefont {Poschinger},\ and\ \citenamefont
  {Schmidt-Kaler}}]{schmiegelow2016transfer}%
  \BibitemOpen
  \bibfield  {author} {\bibinfo {author} {\bibfnamefont {C.~T.}\ \bibnamefont
  {Schmiegelow}}, \bibinfo {author} {\bibfnamefont {J.}~\bibnamefont {Schulz}},
  \bibinfo {author} {\bibfnamefont {H.}~\bibnamefont {Kaufmann}}, \bibinfo
  {author} {\bibfnamefont {T.}~\bibnamefont {Ruster}}, \bibinfo {author}
  {\bibfnamefont {U.~G.}\ \bibnamefont {Poschinger}},\ and\ \bibinfo {author}
  {\bibfnamefont {F.}~\bibnamefont {Schmidt-Kaler}},\ }\bibfield  {title}
  {\bibinfo {title} {Transfer of optical orbital angular momentum to a bound
  electron},\ }\href@noop {} {\bibfield  {journal} {\bibinfo  {journal} {Nat.
  Commun.}\ }\textbf {\bibinfo {volume} {7}},\ \bibinfo {pages} {12998}
  (\bibinfo {year} {2016})}\BibitemShut {NoStop}%
\bibitem [{\citenamefont {Das}\ \emph {et~al.}(2024)\citenamefont {Das},
  \citenamefont {Granados}, \citenamefont {Kr{\"u}ger},\ and\ \citenamefont
  {Ciappina}}]{das2024high}%
  \BibitemOpen
  \bibfield  {author} {\bibinfo {author} {\bibfnamefont {B.~K.}\ \bibnamefont
  {Das}}, \bibinfo {author} {\bibfnamefont {C.}~\bibnamefont {Granados}},
  \bibinfo {author} {\bibfnamefont {M.}~\bibnamefont {Kr{\"u}ger}},\ and\
  \bibinfo {author} {\bibfnamefont {M.~F.}\ \bibnamefont {Ciappina}},\
  }\bibfield  {title} {\bibinfo {title} {High-order harmonic generation driven
  by perfect optical vortex beams: {E}xploring the orbital angular momentum
  upscaling law},\ }\href@noop {} {\bibfield  {journal} {\bibinfo  {journal}
  {Phys. Rev. Research}\ }\textbf {\bibinfo {volume} {6}},\ \bibinfo {pages}
  {043244} (\bibinfo {year} {2024})}\BibitemShut {NoStop}%
\bibitem [{\citenamefont {Shi}\ \emph {et~al.}(2024)\citenamefont {Shi},
  \citenamefont {Zhang}, \citenamefont {Arefiev},\ and\ \citenamefont
  {Shen}}]{shi2024advances}%
  \BibitemOpen
  \bibfield  {author} {\bibinfo {author} {\bibfnamefont {Y.}~\bibnamefont
  {Shi}}, \bibinfo {author} {\bibfnamefont {X.}~\bibnamefont {Zhang}}, \bibinfo
  {author} {\bibfnamefont {A.}~\bibnamefont {Arefiev}},\ and\ \bibinfo {author}
  {\bibfnamefont {B.}~\bibnamefont {Shen}},\ }\bibfield  {title} {\bibinfo
  {title} {Advances in laser-plasma interactions using intense vortex laser
  beams},\ }\href@noop {} {\bibfield  {journal} {\bibinfo  {journal} {Sci.
  China Phys. Mech. Astron.}\ }\textbf {\bibinfo {volume} {67}},\ \bibinfo
  {pages} {295201} (\bibinfo {year} {2024})}\BibitemShut {NoStop}%
\bibitem [{\citenamefont {Forbes}\ and\ \citenamefont
  {Andrews}(2018)}]{forbes2018optical}%
  \BibitemOpen
  \bibfield  {author} {\bibinfo {author} {\bibfnamefont {K.~A.}\ \bibnamefont
  {Forbes}}\ and\ \bibinfo {author} {\bibfnamefont {D.~L.}\ \bibnamefont
  {Andrews}},\ }\bibfield  {title} {\bibinfo {title} {Optical orbital angular
  momentum: twisted light and chirality},\ }\href@noop {} {\bibfield  {journal}
  {\bibinfo  {journal} {Opt. Lett.}\ }\textbf {\bibinfo {volume} {43}},\
  \bibinfo {pages} {435} (\bibinfo {year} {2018})}\BibitemShut {NoStop}%
\bibitem [{\citenamefont {Trawi}\ \emph {et~al.}(2023)\citenamefont {Trawi},
  \citenamefont {Billard}, \citenamefont {Faucher}, \citenamefont {B{\'e}jot},\
  and\ \citenamefont {Hertz}}]{trawi2023molecular}%
  \BibitemOpen
  \bibfield  {author} {\bibinfo {author} {\bibfnamefont {F.}~\bibnamefont
  {Trawi}}, \bibinfo {author} {\bibfnamefont {F.}~\bibnamefont {Billard}},
  \bibinfo {author} {\bibfnamefont {O.}~\bibnamefont {Faucher}}, \bibinfo
  {author} {\bibfnamefont {P.}~\bibnamefont {B{\'e}jot}},\ and\ \bibinfo
  {author} {\bibfnamefont {E.}~\bibnamefont {Hertz}},\ }\bibfield  {title}
  {\bibinfo {title} {Molecular quantum interface for storing and manipulating
  ultrashort optical vortex},\ }\href@noop {} {\bibfield  {journal} {\bibinfo
  {journal} {Laser \& Photonics Reviews}\ }\textbf {\bibinfo {volume} {17}},\
  \bibinfo {pages} {2200525} (\bibinfo {year} {2023})}\BibitemShut {NoStop}%
\bibitem [{\citenamefont {Garc{\'e}s-Ch{\'a}vez}\ \emph
  {et~al.}(2003)\citenamefont {Garc{\'e}s-Ch{\'a}vez}, \citenamefont {McGloin},
  \citenamefont {Padgett}, \citenamefont {Dultz}, \citenamefont {Schmitzer},\
  and\ \citenamefont {Dholakia}}]{garces2003observation}%
  \BibitemOpen
  \bibfield  {author} {\bibinfo {author} {\bibfnamefont {V.}~\bibnamefont
  {Garc{\'e}s-Ch{\'a}vez}}, \bibinfo {author} {\bibfnamefont {D.}~\bibnamefont
  {McGloin}}, \bibinfo {author} {\bibfnamefont {M.}~\bibnamefont {Padgett}},
  \bibinfo {author} {\bibfnamefont {W.}~\bibnamefont {Dultz}}, \bibinfo
  {author} {\bibfnamefont {H.}~\bibnamefont {Schmitzer}},\ and\ \bibinfo
  {author} {\bibfnamefont {K.}~\bibnamefont {Dholakia}},\ }\bibfield  {title}
  {\bibinfo {title} {Observation of the {T}ransfer of the {L}ocal {A}ngular
  {M}omentum {D}ensity of a {M}ultiringed {L}ight {B}eam to an {O}ptically
  {T}rapped {P}article},\ }\href@noop {} {\bibfield  {journal} {\bibinfo
  {journal} {Phys. Rev. Lett.}\ }\textbf {\bibinfo {volume} {91}},\ \bibinfo
  {pages} {093602} (\bibinfo {year} {2003})}\BibitemShut {NoStop}%
\bibitem [{\citenamefont {Swartzlander~Jr}\ \emph {et~al.}(2008)\citenamefont
  {Swartzlander~Jr}, \citenamefont {Ford}, \citenamefont {Abdul-Malik},
  \citenamefont {Close}, \citenamefont {Peters}, \citenamefont {Palacios},\
  and\ \citenamefont {Wilson}}]{swartzlander2008astronomical}%
  \BibitemOpen
  \bibfield  {author} {\bibinfo {author} {\bibfnamefont {G.~A.}\ \bibnamefont
  {Swartzlander~Jr}}, \bibinfo {author} {\bibfnamefont {E.~L.}\ \bibnamefont
  {Ford}}, \bibinfo {author} {\bibfnamefont {R.~S.}\ \bibnamefont
  {Abdul-Malik}}, \bibinfo {author} {\bibfnamefont {L.~M.}\ \bibnamefont
  {Close}}, \bibinfo {author} {\bibfnamefont {M.~A.}\ \bibnamefont {Peters}},
  \bibinfo {author} {\bibfnamefont {D.~M.}\ \bibnamefont {Palacios}},\ and\
  \bibinfo {author} {\bibfnamefont {D.~W.}\ \bibnamefont {Wilson}},\ }\bibfield
   {title} {\bibinfo {title} {Astronomical demonstration of an optical vortex
  coronagraph},\ }\href@noop {} {\bibfield  {journal} {\bibinfo  {journal}
  {Opt. Express}\ }\textbf {\bibinfo {volume} {16}},\ \bibinfo {pages} {10200}
  (\bibinfo {year} {2008})}\BibitemShut {NoStop}%
\bibitem [{\citenamefont {Lu}\ \emph {et~al.}(2023)\citenamefont {Lu},
  \citenamefont {Guo}, \citenamefont {Li}, \citenamefont {Ababekri},
  \citenamefont {Chen}, \citenamefont {Fu}, \citenamefont {Lv}, \citenamefont
  {Xu}, \citenamefont {Kong}, \citenamefont {Niu},\ and\ \citenamefont
  {Li}}]{lu2023manipulation}%
  \BibitemOpen
  \bibfield  {author} {\bibinfo {author} {\bibfnamefont {Z.-W.}\ \bibnamefont
  {Lu}}, \bibinfo {author} {\bibfnamefont {L.}~\bibnamefont {Guo}}, \bibinfo
  {author} {\bibfnamefont {Z.-Z.}\ \bibnamefont {Li}}, \bibinfo {author}
  {\bibfnamefont {M.}~\bibnamefont {Ababekri}}, \bibinfo {author}
  {\bibfnamefont {F.-Q.}\ \bibnamefont {Chen}}, \bibinfo {author}
  {\bibfnamefont {C.}~\bibnamefont {Fu}}, \bibinfo {author} {\bibfnamefont
  {C.}~\bibnamefont {Lv}}, \bibinfo {author} {\bibfnamefont {R.}~\bibnamefont
  {Xu}}, \bibinfo {author} {\bibfnamefont {X.}~\bibnamefont {Kong}}, \bibinfo
  {author} {\bibfnamefont {Y.-F.}\ \bibnamefont {Niu}},\ and\ \bibinfo {author}
  {\bibfnamefont {J.-X.}\ \bibnamefont {Li}},\ }\bibfield  {title} {\bibinfo
  {title} {Manipulation of giant multipole resonances via vortex $\gamma$
  photons},\ }\href@noop {} {\bibfield  {journal} {\bibinfo  {journal} {Phys.
  Rev. Lett.}\ }\textbf {\bibinfo {volume} {131}},\ \bibinfo {pages} {202502}
  (\bibinfo {year} {2023})}\BibitemShut {NoStop}%
\bibitem [{\citenamefont {Lu}\ \emph {et~al.}(2025)\citenamefont {Lu},
  \citenamefont {Guo}, \citenamefont {Ababekri}, \citenamefont {Zhang},
  \citenamefont {Weng}, \citenamefont {Wu}, \citenamefont {Niu},\ and\
  \citenamefont {Li}}]{lu2025angular}%
  \BibitemOpen
  \bibfield  {author} {\bibinfo {author} {\bibfnamefont {Z.-W.}\ \bibnamefont
  {Lu}}, \bibinfo {author} {\bibfnamefont {L.}~\bibnamefont {Guo}}, \bibinfo
  {author} {\bibfnamefont {M.}~\bibnamefont {Ababekri}}, \bibinfo {author}
  {\bibfnamefont {J.-L.}\ \bibnamefont {Zhang}}, \bibinfo {author}
  {\bibfnamefont {X.-F.}\ \bibnamefont {Weng}}, \bibinfo {author}
  {\bibfnamefont {Y.}~\bibnamefont {Wu}}, \bibinfo {author} {\bibfnamefont
  {Y.-F.}\ \bibnamefont {Niu}},\ and\ \bibinfo {author} {\bibfnamefont {J.-X.}\
  \bibnamefont {Li}},\ }\bibfield  {title} {\bibinfo {title} {{A}ngular
  {M}omentum {R}esolved {I}nelastic {E}lectron {S}cattering for {N}uclear
  {G}iant {R}esonances},\ }\href@noop {} {\bibfield  {journal} {\bibinfo
  {journal} {Phys. Rev. Lett.}\ }\textbf {\bibinfo {volume} {134}},\ \bibinfo
  {pages} {052501} (\bibinfo {year} {2025})}\BibitemShut {NoStop}%
\bibitem [{\citenamefont {Bohr}(1976)}]{bohr1976rotational}%
  \BibitemOpen
  \bibfield  {author} {\bibinfo {author} {\bibfnamefont {A.}~\bibnamefont
  {Bohr}},\ }\bibfield  {title} {\bibinfo {title} {Rotational motion in
  nuclei},\ }\href@noop {} {\bibfield  {journal} {\bibinfo  {journal} {Rev.
  Mod. Phys.}\ }\textbf {\bibinfo {volume} {48}},\ \bibinfo {pages} {365}
  (\bibinfo {year} {1976})}\BibitemShut {NoStop}%
\bibitem [{\citenamefont {Eisenberg}\ and\ \citenamefont
  {Greiner}(1976)}]{eisenberg1976nuclear}%
  \BibitemOpen
  \bibfield  {author} {\bibinfo {author} {\bibfnamefont {J.~M.}\ \bibnamefont
  {Eisenberg}}\ and\ \bibinfo {author} {\bibfnamefont {W.}~\bibnamefont
  {Greiner}},\ }\bibfield  {title} {\bibinfo {title} {Nuclear {T}heory.
  {E}xcitation mechanisms of the nucleus. vol. 2},\ }\href@noop {} {\
  (\bibinfo {year} {1976})}\BibitemShut {NoStop}%
\bibitem [{\citenamefont {Kirschbaum}\ \emph {et~al.}(2024)\citenamefont
  {Kirschbaum}, \citenamefont {Schumm},\ and\ \citenamefont
  {P{\'a}lffy}}]{kirschbaum2024photoexcitation}%
  \BibitemOpen
  \bibfield  {author} {\bibinfo {author} {\bibfnamefont {T.}~\bibnamefont
  {Kirschbaum}}, \bibinfo {author} {\bibfnamefont {T.}~\bibnamefont {Schumm}},\
  and\ \bibinfo {author} {\bibfnamefont {A.}~\bibnamefont {P{\'a}lffy}},\
  }\bibfield  {title} {\bibinfo {title} {Photoexcitation of the $^{229}${T}h
  nuclear clock transition using twisted light},\ }\href@noop {} {\bibfield
  {journal} {\bibinfo  {journal} {Phys. Rev. C}\ }\textbf {\bibinfo {volume}
  {110}},\ \bibinfo {pages} {064326} (\bibinfo {year} {2024})}\BibitemShut
  {NoStop}%
\bibitem [{\citenamefont {Alder}\ \emph {et~al.}(1956)\citenamefont {Alder},
  \citenamefont {Bohr}, \citenamefont {Huus}, \citenamefont {Mottelson},\ and\
  \citenamefont {Winther}}]{alder1956study}%
  \BibitemOpen
  \bibfield  {author} {\bibinfo {author} {\bibfnamefont {K.}~\bibnamefont
  {Alder}}, \bibinfo {author} {\bibfnamefont {A.}~\bibnamefont {Bohr}},
  \bibinfo {author} {\bibfnamefont {T.}~\bibnamefont {Huus}}, \bibinfo {author}
  {\bibfnamefont {B.}~\bibnamefont {Mottelson}},\ and\ \bibinfo {author}
  {\bibfnamefont {A.}~\bibnamefont {Winther}},\ }\bibfield  {title} {\bibinfo
  {title} {Study of nuclear structure by electromagnetic excitation with
  accelerated ions},\ }\href@noop {} {\bibfield  {journal} {\bibinfo  {journal}
  {Rev. Mod. Phys.}\ }\textbf {\bibinfo {volume} {28}},\ \bibinfo {pages} {432}
  (\bibinfo {year} {1956})}\BibitemShut {NoStop}%
\bibitem [{\citenamefont {Cowan}(1981)}]{cowan1981theory}%
  \BibitemOpen
  \bibfield  {author} {\bibinfo {author} {\bibfnamefont {R.~D.}\ \bibnamefont
  {Cowan}},\ }\href@noop {} {\emph {\bibinfo {title} {The {T}heory of {A}tomic
  {S}tructure and {S}pectra}}},\ \bibinfo {number} {3}\ (\bibinfo  {publisher}
  {University of California Press, Berkerley},\ \bibinfo {year}
  {1981})\BibitemShut {NoStop}%
\bibitem [{sup()}]{supplemental}%
  \BibitemOpen
  \href@noop {} {}\bibinfo {note} {See {S}upplemental {M}aterial for more
  details on the derivations of the NHM effect and vortex laser–nucleus
  interaction, as well as the comprehensive analysis of the results presented
  in this paper.}\BibitemShut {Stop}%
\bibitem [{\citenamefont {Sheppard}\ and\ \citenamefont
  {Wilson}(1978)}]{sheppard1978gaussian}%
  \BibitemOpen
  \bibfield  {author} {\bibinfo {author} {\bibfnamefont {C.}~\bibnamefont
  {Sheppard}}\ and\ \bibinfo {author} {\bibfnamefont {T.}~\bibnamefont
  {Wilson}},\ }\bibfield  {title} {\bibinfo {title} {Gaussian-beam theory of
  lenses with annular aperture},\ }\href@noop {} {\bibfield  {journal}
  {\bibinfo  {journal} {IEEE J. Microw. Opt. Acoust.}\ }\textbf {\bibinfo
  {volume} {2}},\ \bibinfo {pages} {105} (\bibinfo {year} {1978})}\BibitemShut
  {NoStop}%
\bibitem [{\citenamefont {Durnin}(1987)}]{durnin1987exact}%
  \BibitemOpen
  \bibfield  {author} {\bibinfo {author} {\bibfnamefont {J.}~\bibnamefont
  {Durnin}},\ }\bibfield  {title} {\bibinfo {title} {Exact solutions for
  nondiffracting beams. i. the scalar theory},\ }\href@noop {} {\bibfield
  {journal} {\bibinfo  {journal} {J. Opt. Soc. Am.A}\ }\textbf {\bibinfo
  {volume} {4}},\ \bibinfo {pages} {651} (\bibinfo {year} {1987})}\BibitemShut
  {NoStop}%
\bibitem [{\citenamefont {Schulz}\ \emph {et~al.}(2020)\citenamefont {Schulz},
  \citenamefont {Peshkov}, \citenamefont {M{\"u}ller}, \citenamefont {Lange},
  \citenamefont {Huntemann}, \citenamefont {Tamm}, \citenamefont {Peik},\ and\
  \citenamefont {Surzhykov}}]{schulz2020generalized}%
  \BibitemOpen
  \bibfield  {author} {\bibinfo {author} {\bibfnamefont {S.-L.}\ \bibnamefont
  {Schulz}}, \bibinfo {author} {\bibfnamefont {A.}~\bibnamefont {Peshkov}},
  \bibinfo {author} {\bibfnamefont {R.}~\bibnamefont {M{\"u}ller}}, \bibinfo
  {author} {\bibfnamefont {R.}~\bibnamefont {Lange}}, \bibinfo {author}
  {\bibfnamefont {N.}~\bibnamefont {Huntemann}}, \bibinfo {author}
  {\bibfnamefont {C.}~\bibnamefont {Tamm}}, \bibinfo {author} {\bibfnamefont
  {E.}~\bibnamefont {Peik}},\ and\ \bibinfo {author} {\bibfnamefont
  {A.}~\bibnamefont {Surzhykov}},\ }\bibfield  {title} {\bibinfo {title}
  {Generalized excitation of atomic multipole transitions by twisted light
  modes},\ }\href@noop {} {\bibfield  {journal} {\bibinfo  {journal} {Phys.
  Rev. A}\ }\textbf {\bibinfo {volume} {102}},\ \bibinfo {pages} {012812}
  (\bibinfo {year} {2020})}\BibitemShut {NoStop}%
\bibitem [{\citenamefont {Safronova}\ \emph {et~al.}(2013)\citenamefont
  {Safronova}, \citenamefont {Safronova}, \citenamefont {Radnaev},
  \citenamefont {Campbell},\ and\ \citenamefont
  {Kuzmich}}]{safronova2013magnetic}%
  \BibitemOpen
  \bibfield  {author} {\bibinfo {author} {\bibfnamefont {M.}~\bibnamefont
  {Safronova}}, \bibinfo {author} {\bibfnamefont {U.}~\bibnamefont
  {Safronova}}, \bibinfo {author} {\bibfnamefont {A.}~\bibnamefont {Radnaev}},
  \bibinfo {author} {\bibfnamefont {C.}~\bibnamefont {Campbell}},\ and\
  \bibinfo {author} {\bibfnamefont {A.}~\bibnamefont {Kuzmich}},\ }\bibfield
  {title} {\bibinfo {title} {Magnetic dipole and electric quadrupole moments of
  the $^{229}${T}h nucleus},\ }\href@noop {} {\bibfield  {journal} {\bibinfo
  {journal} {Phys. Rev. A}\ }\textbf {\bibinfo {volume} {88}},\ \bibinfo
  {pages} {060501} (\bibinfo {year} {2013})}\BibitemShut {NoStop}%
\bibitem [{\citenamefont {Thielking}\ \emph {et~al.}(2018)\citenamefont
  {Thielking}, \citenamefont {Okhapkin}, \citenamefont {G{\l}owacki},
  \citenamefont {Meier}, \citenamefont {von~der Wense}, \citenamefont
  {Seiferle}, \citenamefont {D{\"u}llmann}, \citenamefont {Thirolf},\ and\
  \citenamefont {Peik}}]{thielking2018laser}%
  \BibitemOpen
  \bibfield  {author} {\bibinfo {author} {\bibfnamefont {J.}~\bibnamefont
  {Thielking}}, \bibinfo {author} {\bibfnamefont {M.~V.}\ \bibnamefont
  {Okhapkin}}, \bibinfo {author} {\bibfnamefont {P.}~\bibnamefont
  {G{\l}owacki}}, \bibinfo {author} {\bibfnamefont {D.~M.}\ \bibnamefont
  {Meier}}, \bibinfo {author} {\bibfnamefont {L.}~\bibnamefont {von~der
  Wense}}, \bibinfo {author} {\bibfnamefont {B.}~\bibnamefont {Seiferle}},
  \bibinfo {author} {\bibfnamefont {C.~E.}\ \bibnamefont {D{\"u}llmann}},
  \bibinfo {author} {\bibfnamefont {P.~G.}\ \bibnamefont {Thirolf}},\ and\
  \bibinfo {author} {\bibfnamefont {E.}~\bibnamefont {Peik}},\ }\bibfield
  {title} {\bibinfo {title} {Laser spectroscopic characterization of the
  nuclear-clock isomer $^{229m}${T}h},\ }\href@noop {} {\bibfield  {journal}
  {\bibinfo  {journal} {Nature}\ }\textbf {\bibinfo {volume} {556}},\ \bibinfo
  {pages} {321} (\bibinfo {year} {2018})}\BibitemShut {NoStop}%
\bibitem [{\citenamefont {Minkov}\ and\ \citenamefont
  {P{\'a}lffy}(2021)}]{minkov2021th}%
  \BibitemOpen
  \bibfield  {author} {\bibinfo {author} {\bibfnamefont {N.}~\bibnamefont
  {Minkov}}\ and\ \bibinfo {author} {\bibfnamefont {A.}~\bibnamefont
  {P{\'a}lffy}},\ }\bibfield  {title} {\bibinfo {title} {$^{229m}${T}h isomer
  from a nuclear model perspective},\ }\href@noop {} {\bibfield  {journal}
  {\bibinfo  {journal} {Phys. Rev. C}\ }\textbf {\bibinfo {volume} {103}},\
  \bibinfo {pages} {014313} (\bibinfo {year} {2021})}\BibitemShut {NoStop}%
\bibitem [{\citenamefont {Verner}(2010)}]{verner2010numerically}%
  \BibitemOpen
  \bibfield  {author} {\bibinfo {author} {\bibfnamefont {J.~H.}\ \bibnamefont
  {Verner}},\ }\bibfield  {title} {\bibinfo {title} {Numerically optimal
  {R}unge--{K}utta pairs with interpolants},\ }\href@noop {} {\bibfield
  {journal} {\bibinfo  {journal} {Numer. Algorithms}\ }\textbf {\bibinfo
  {volume} {53}},\ \bibinfo {pages} {383} (\bibinfo {year} {2010})}\BibitemShut
  {NoStop}%
\bibitem [{\citenamefont {Greiner}(2011)}]{greiner2011quantum}%
  \BibitemOpen
  \bibfield  {author} {\bibinfo {author} {\bibfnamefont {W.}~\bibnamefont
  {Greiner}},\ }\href@noop {} {\emph {\bibinfo {title} {Quantum mechanics: an
  introduction}}}\ (\bibinfo  {publisher} {Springer Science \& Business
  Media},\ \bibinfo {year} {2011})\BibitemShut {NoStop}%
\bibitem [{\citenamefont {Yuan}\ \emph {et~al.}(2022)\citenamefont {Yuan},
  \citenamefont {Wang}, \citenamefont {Li}, \citenamefont {Wang}, \citenamefont
  {Wang}, \citenamefont {Huang}, \citenamefont {Li}, \citenamefont {Ma},
  \citenamefont {Zhu}, \citenamefont {Xu} \emph {et~al.}}]{yuan2022nuclear}%
  \BibitemOpen
  \bibfield  {author} {\bibinfo {author} {\bibfnamefont {Z.}~\bibnamefont
  {Yuan}}, \bibinfo {author} {\bibfnamefont {H.}~\bibnamefont {Wang}}, \bibinfo
  {author} {\bibfnamefont {Z.}~\bibnamefont {Li}}, \bibinfo {author}
  {\bibfnamefont {T.}~\bibnamefont {Wang}}, \bibinfo {author} {\bibfnamefont
  {H.}~\bibnamefont {Wang}}, \bibinfo {author} {\bibfnamefont {X.}~\bibnamefont
  {Huang}}, \bibinfo {author} {\bibfnamefont {T.}~\bibnamefont {Li}}, \bibinfo
  {author} {\bibfnamefont {Z.}~\bibnamefont {Ma}}, \bibinfo {author}
  {\bibfnamefont {L.}~\bibnamefont {Zhu}}, \bibinfo {author} {\bibfnamefont
  {W.}~\bibnamefont {Xu}}, \emph {et~al.},\ }\bibfield  {title} {\bibinfo
  {title} {Nuclear phase retrieval spectroscopy using resonant x-ray
  scattering},\ }\href@noop {} {\bibfield  {journal} {\bibinfo  {journal}
  {arXiv:2204.06096}\ } (\bibinfo {year} {2022})}\BibitemShut {NoStop}%
\bibitem [{\citenamefont {Steck}\ and\ \citenamefont
  {Litvinov}(2020)}]{steck2020heavy}%
  \BibitemOpen
  \bibfield  {author} {\bibinfo {author} {\bibfnamefont {M.}~\bibnamefont
  {Steck}}\ and\ \bibinfo {author} {\bibfnamefont {Y.~A.}\ \bibnamefont
  {Litvinov}},\ }\bibfield  {title} {\bibinfo {title} {Heavy-ion storage rings
  and their use in precision experiments with highly charged ions},\
  }\href@noop {} {\bibfield  {journal} {\bibinfo  {journal} {Prog. Part. Nucl.
  Phys.}\ }\textbf {\bibinfo {volume} {115}},\ \bibinfo {pages} {103811}
  (\bibinfo {year} {2020})}\BibitemShut {NoStop}%
\bibitem [{\citenamefont {Ma}\ \emph {et~al.}(2015)\citenamefont {Ma},
  \citenamefont {Wen}, \citenamefont {Huang}, \citenamefont {Wang},
  \citenamefont {Yuan}, \citenamefont {Wang}, \citenamefont {Sun},
  \citenamefont {Mao}, \citenamefont {Yang}, \citenamefont {Xu} \emph
  {et~al.}}]{ma2015proposal}%
  \BibitemOpen
  \bibfield  {author} {\bibinfo {author} {\bibfnamefont {X.}~\bibnamefont
  {Ma}}, \bibinfo {author} {\bibfnamefont {W.}~\bibnamefont {Wen}}, \bibinfo
  {author} {\bibfnamefont {Z.}~\bibnamefont {Huang}}, \bibinfo {author}
  {\bibfnamefont {H.}~\bibnamefont {Wang}}, \bibinfo {author} {\bibfnamefont
  {Y.}~\bibnamefont {Yuan}}, \bibinfo {author} {\bibfnamefont {M.}~\bibnamefont
  {Wang}}, \bibinfo {author} {\bibfnamefont {Z.}~\bibnamefont {Sun}}, \bibinfo
  {author} {\bibfnamefont {L.}~\bibnamefont {Mao}}, \bibinfo {author}
  {\bibfnamefont {J.}~\bibnamefont {Yang}}, \bibinfo {author} {\bibfnamefont
  {H.}~\bibnamefont {Xu}}, \emph {et~al.},\ }\bibfield  {title} {\bibinfo
  {title} {Proposal for precision determination of 7.8 e{V} isomeric state in
  $^{229}${T}h at heavy ion storage ring},\ }\href@noop {} {\bibfield
  {journal} {\bibinfo  {journal} {Phys. Scr.}\ }\textbf {\bibinfo {volume}
  {2015}},\ \bibinfo {pages} {014012} (\bibinfo {year} {2015})}\BibitemShut
  {NoStop}%
\bibitem [{\citenamefont {Ringleb}\ \emph {et~al.}(2022)\citenamefont
  {Ringleb}, \citenamefont {Kiffer}, \citenamefont {Stallkamp}, \citenamefont
  {Kumar}, \citenamefont {Hofbrucker}, \citenamefont {Reich}, \citenamefont
  {Arndt}, \citenamefont {Brenner}, \citenamefont {Ruiz-Lop{\'e}z},
  \citenamefont {D{\"u}sterer} \emph {et~al.}}]{ringleb2022high}%
  \BibitemOpen
  \bibfield  {author} {\bibinfo {author} {\bibfnamefont {S.}~\bibnamefont
  {Ringleb}}, \bibinfo {author} {\bibfnamefont {M.}~\bibnamefont {Kiffer}},
  \bibinfo {author} {\bibfnamefont {N.}~\bibnamefont {Stallkamp}}, \bibinfo
  {author} {\bibfnamefont {S.}~\bibnamefont {Kumar}}, \bibinfo {author}
  {\bibfnamefont {J.}~\bibnamefont {Hofbrucker}}, \bibinfo {author}
  {\bibfnamefont {B.}~\bibnamefont {Reich}}, \bibinfo {author} {\bibfnamefont
  {B.}~\bibnamefont {Arndt}}, \bibinfo {author} {\bibfnamefont
  {G.}~\bibnamefont {Brenner}}, \bibinfo {author} {\bibfnamefont
  {M.}~\bibnamefont {Ruiz-Lop{\'e}z}}, \bibinfo {author} {\bibfnamefont
  {S.}~\bibnamefont {D{\"u}sterer}}, \emph {et~al.},\ }\bibfield  {title}
  {\bibinfo {title} {High-intensity laser experiments with highly charged ions
  in a {P}enning trap},\ }\href@noop {} {\bibfield  {journal} {\bibinfo
  {journal} {Phys. Scr.}\ }\textbf {\bibinfo {volume} {97}},\ \bibinfo {pages}
  {084002} (\bibinfo {year} {2022})}\BibitemShut {NoStop}%
\bibitem [{\citenamefont {Major}\ \emph {et~al.}(2024)\citenamefont {Major},
  \citenamefont {Eisenbarth}, \citenamefont {Zielbauer}, \citenamefont
  {Brabetz}, \citenamefont {Ohland}, \citenamefont {Zobus}, \citenamefont
  {Roeder}, \citenamefont {Reemts}, \citenamefont {Kunzer}, \citenamefont
  {G{\"o}tte} \emph {et~al.}}]{major2024high}%
  \BibitemOpen
  \bibfield  {author} {\bibinfo {author} {\bibfnamefont {Z.}~\bibnamefont
  {Major}}, \bibinfo {author} {\bibfnamefont {U.}~\bibnamefont {Eisenbarth}},
  \bibinfo {author} {\bibfnamefont {B.}~\bibnamefont {Zielbauer}}, \bibinfo
  {author} {\bibfnamefont {C.}~\bibnamefont {Brabetz}}, \bibinfo {author}
  {\bibfnamefont {J.}~\bibnamefont {Ohland}}, \bibinfo {author} {\bibfnamefont
  {Y.}~\bibnamefont {Zobus}}, \bibinfo {author} {\bibfnamefont
  {S.}~\bibnamefont {Roeder}}, \bibinfo {author} {\bibfnamefont
  {D.}~\bibnamefont {Reemts}}, \bibinfo {author} {\bibfnamefont
  {S.}~\bibnamefont {Kunzer}}, \bibinfo {author} {\bibfnamefont
  {S.}~\bibnamefont {G{\"o}tte}}, \emph {et~al.},\ }\bibfield  {title}
  {\bibinfo {title} {High-energy laser facility {PHELIX} at {GSI}: latest
  advances and extended capabilities},\ }\href@noop {} {\bibfield  {journal}
  {\bibinfo  {journal} {High Power Laser Sci. Eng.}\ }\textbf {\bibinfo
  {volume} {12}},\ \bibinfo {pages} {e39} (\bibinfo {year} {2024})}\BibitemShut
  {NoStop}%
\bibitem [{\citenamefont {Bernitt}\ \emph {et~al.}(2012)\citenamefont
  {Bernitt}, \citenamefont {Brown}, \citenamefont {Rudolph}, \citenamefont
  {Steinbr{\"u}gge}, \citenamefont {Graf}, \citenamefont {Leutenegger},
  \citenamefont {Epp}, \citenamefont {Eberle}, \citenamefont {Kubi{\v{c}}ek},
  \citenamefont {M{\"a}ckel} \emph {et~al.}}]{bernitt2012unexpectedly}%
  \BibitemOpen
  \bibfield  {author} {\bibinfo {author} {\bibfnamefont {S.}~\bibnamefont
  {Bernitt}}, \bibinfo {author} {\bibfnamefont {G.}~\bibnamefont {Brown}},
  \bibinfo {author} {\bibfnamefont {J.~K.}\ \bibnamefont {Rudolph}}, \bibinfo
  {author} {\bibfnamefont {R.}~\bibnamefont {Steinbr{\"u}gge}}, \bibinfo
  {author} {\bibfnamefont {A.}~\bibnamefont {Graf}}, \bibinfo {author}
  {\bibfnamefont {M.}~\bibnamefont {Leutenegger}}, \bibinfo {author}
  {\bibfnamefont {S.}~\bibnamefont {Epp}}, \bibinfo {author} {\bibfnamefont
  {S.}~\bibnamefont {Eberle}}, \bibinfo {author} {\bibfnamefont
  {K.}~\bibnamefont {Kubi{\v{c}}ek}}, \bibinfo {author} {\bibfnamefont
  {V.}~\bibnamefont {M{\"a}ckel}}, \emph {et~al.},\ }\bibfield  {title}
  {\bibinfo {title} {An unexpectedly low oscillator strength as the origin of
  the {F}e {XVII} emission problem},\ }\href@noop {} {\bibfield  {journal}
  {\bibinfo  {journal} {Nature}\ }\textbf {\bibinfo {volume} {492}},\ \bibinfo
  {pages} {225} (\bibinfo {year} {2012})}\BibitemShut {NoStop}%
\bibitem [{\citenamefont {Sturm}\ \emph {et~al.}(2019)\citenamefont {Sturm},
  \citenamefont {Arapoglou}, \citenamefont {Egl}, \citenamefont {H{\"o}cker},
  \citenamefont {Kraemer}, \citenamefont {Sailer}, \citenamefont {Tu},
  \citenamefont {Weigel}, \citenamefont {Wolf}, \citenamefont
  {L{\'o}pez-Urrutia} \emph {et~al.}}]{sturm2019alphatrap}%
  \BibitemOpen
  \bibfield  {author} {\bibinfo {author} {\bibfnamefont {S.}~\bibnamefont
  {Sturm}}, \bibinfo {author} {\bibfnamefont {I.}~\bibnamefont {Arapoglou}},
  \bibinfo {author} {\bibfnamefont {A.}~\bibnamefont {Egl}}, \bibinfo {author}
  {\bibfnamefont {M.}~\bibnamefont {H{\"o}cker}}, \bibinfo {author}
  {\bibfnamefont {S.}~\bibnamefont {Kraemer}}, \bibinfo {author} {\bibfnamefont
  {T.}~\bibnamefont {Sailer}}, \bibinfo {author} {\bibfnamefont
  {B.}~\bibnamefont {Tu}}, \bibinfo {author} {\bibfnamefont {A.}~\bibnamefont
  {Weigel}}, \bibinfo {author} {\bibfnamefont {R.}~\bibnamefont {Wolf}},
  \bibinfo {author} {\bibfnamefont {J.~C.}\ \bibnamefont {L{\'o}pez-Urrutia}},
  \emph {et~al.},\ }\bibfield  {title} {\bibinfo {title} {The {ALPHATRAP}
  experiment},\ }\href@noop {} {\bibfield  {journal} {\bibinfo  {journal} {Eur.
  Phys. J. Spec. Top.}\ }\textbf {\bibinfo {volume} {227}},\ \bibinfo {pages}
  {1425} (\bibinfo {year} {2019})}\BibitemShut {NoStop}%
\bibitem [{\citenamefont {Brabetz}\ \emph {et~al.}(2015)\citenamefont
  {Brabetz}, \citenamefont {Busold}, \citenamefont {Cowan}, \citenamefont
  {Deppert}, \citenamefont {Jahn}, \citenamefont {Kester}, \citenamefont
  {Roth}, \citenamefont {Schumacher},\ and\ \citenamefont
  {Bagnoud}}]{brabetz2015laser}%
  \BibitemOpen
  \bibfield  {author} {\bibinfo {author} {\bibfnamefont {C.}~\bibnamefont
  {Brabetz}}, \bibinfo {author} {\bibfnamefont {S.}~\bibnamefont {Busold}},
  \bibinfo {author} {\bibfnamefont {T.}~\bibnamefont {Cowan}}, \bibinfo
  {author} {\bibfnamefont {O.}~\bibnamefont {Deppert}}, \bibinfo {author}
  {\bibfnamefont {D.}~\bibnamefont {Jahn}}, \bibinfo {author} {\bibfnamefont
  {O.}~\bibnamefont {Kester}}, \bibinfo {author} {\bibfnamefont
  {M.}~\bibnamefont {Roth}}, \bibinfo {author} {\bibfnamefont {D.}~\bibnamefont
  {Schumacher}},\ and\ \bibinfo {author} {\bibfnamefont {V.}~\bibnamefont
  {Bagnoud}},\ }\bibfield  {title} {\bibinfo {title} {Laser-driven ion
  acceleration with hollow laser beams},\ }\href@noop {} {\bibfield  {journal}
  {\bibinfo  {journal} {Phys. Plasmas}\ }\textbf {\bibinfo {volume} {22}}
  (\bibinfo {year} {2015})}\BibitemShut {NoStop}%
\bibitem [{\citenamefont {Wang}\ \emph {et~al.}(2020)\citenamefont {Wang},
  \citenamefont {Jiang}, \citenamefont {Dong}, \citenamefont {Lu},
  \citenamefont {Li}, \citenamefont {Xu}, \citenamefont {Sun}, \citenamefont
  {Yu}, \citenamefont {Guo}, \citenamefont {Liang} \emph
  {et~al.}}]{wang2020hollow}%
  \BibitemOpen
  \bibfield  {author} {\bibinfo {author} {\bibfnamefont {W.}~\bibnamefont
  {Wang}}, \bibinfo {author} {\bibfnamefont {C.}~\bibnamefont {Jiang}},
  \bibinfo {author} {\bibfnamefont {H.}~\bibnamefont {Dong}}, \bibinfo {author}
  {\bibfnamefont {X.}~\bibnamefont {Lu}}, \bibinfo {author} {\bibfnamefont
  {J.}~\bibnamefont {Li}}, \bibinfo {author} {\bibfnamefont {R.}~\bibnamefont
  {Xu}}, \bibinfo {author} {\bibfnamefont {Y.}~\bibnamefont {Sun}}, \bibinfo
  {author} {\bibfnamefont {L.}~\bibnamefont {Yu}}, \bibinfo {author}
  {\bibfnamefont {Z.}~\bibnamefont {Guo}}, \bibinfo {author} {\bibfnamefont
  {X.}~\bibnamefont {Liang}}, \emph {et~al.},\ }\bibfield  {title} {\bibinfo
  {title} {Hollow plasma acceleration driven by a relativistic reflected hollow
  laser},\ }\href@noop {} {\bibfield  {journal} {\bibinfo  {journal} {Phys.
  Rev. Lett.}\ }\textbf {\bibinfo {volume} {125}},\ \bibinfo {pages} {034801}
  (\bibinfo {year} {2020})}\BibitemShut {NoStop}%
\bibitem [{\citenamefont {Radier}\ \emph {et~al.}(2022)\citenamefont {Radier},
  \citenamefont {Chalus}, \citenamefont {Charbonneau}, \citenamefont
  {Thambirajah}, \citenamefont {Deschamps}, \citenamefont {David},
  \citenamefont {Barbe}, \citenamefont {Etter}, \citenamefont {Matras},
  \citenamefont {Ricaud} \emph {et~al.}}]{radier202210}%
  \BibitemOpen
  \bibfield  {author} {\bibinfo {author} {\bibfnamefont {C.}~\bibnamefont
  {Radier}}, \bibinfo {author} {\bibfnamefont {O.}~\bibnamefont {Chalus}},
  \bibinfo {author} {\bibfnamefont {M.}~\bibnamefont {Charbonneau}}, \bibinfo
  {author} {\bibfnamefont {S.}~\bibnamefont {Thambirajah}}, \bibinfo {author}
  {\bibfnamefont {G.}~\bibnamefont {Deschamps}}, \bibinfo {author}
  {\bibfnamefont {S.}~\bibnamefont {David}}, \bibinfo {author} {\bibfnamefont
  {J.}~\bibnamefont {Barbe}}, \bibinfo {author} {\bibfnamefont
  {E.}~\bibnamefont {Etter}}, \bibinfo {author} {\bibfnamefont
  {G.}~\bibnamefont {Matras}}, \bibinfo {author} {\bibfnamefont
  {S.}~\bibnamefont {Ricaud}}, \emph {et~al.},\ }\bibfield  {title} {\bibinfo
  {title} {10 {PW} peak power femtosecond laser pulses at {ELI-NP}},\
  }\href@noop {} {\bibfield  {journal} {\bibinfo  {journal} {High Power Laser
  Sci. Eng.}\ }\textbf {\bibinfo {volume} {10}},\ \bibinfo {pages} {e21}
  (\bibinfo {year} {2022})}\BibitemShut {NoStop}%
\bibitem [{\citenamefont {Wang}\ \emph {et~al.}(2022)\citenamefont {Wang},
  \citenamefont {Liu}, \citenamefont {Lu}, \citenamefont {Chen}, \citenamefont
  {Long}, \citenamefont {Li}, \citenamefont {Chen}, \citenamefont {Chen},
  \citenamefont {Bai}, \citenamefont {Li} \emph {et~al.}}]{wang202213}%
  \BibitemOpen
  \bibfield  {author} {\bibinfo {author} {\bibfnamefont {X.}~\bibnamefont
  {Wang}}, \bibinfo {author} {\bibfnamefont {X.}~\bibnamefont {Liu}}, \bibinfo
  {author} {\bibfnamefont {X.}~\bibnamefont {Lu}}, \bibinfo {author}
  {\bibfnamefont {J.}~\bibnamefont {Chen}}, \bibinfo {author} {\bibfnamefont
  {Y.}~\bibnamefont {Long}}, \bibinfo {author} {\bibfnamefont {W.}~\bibnamefont
  {Li}}, \bibinfo {author} {\bibfnamefont {H.}~\bibnamefont {Chen}}, \bibinfo
  {author} {\bibfnamefont {X.}~\bibnamefont {Chen}}, \bibinfo {author}
  {\bibfnamefont {P.}~\bibnamefont {Bai}}, \bibinfo {author} {\bibfnamefont
  {Y.}~\bibnamefont {Li}}, \emph {et~al.},\ }\bibfield  {title} {\bibinfo
  {title} {13.4 fs, 0.1 {H}z {OPCPA} front end for the 100 {PW}-class laser
  facility},\ }\href@noop {} {\bibfield  {journal} {\bibinfo  {journal}
  {Ultrafast Sci.}\ } (\bibinfo {year} {2022})}\BibitemShut {NoStop}%
\end{thebibliography}%

\end{document}